%% file: MN-19-0699-MJ_final.tex
\documentclass[fleqn,usenatbib]{mnras}

\usepackage{newtxtext,newtxmath}

\usepackage[T1]{fontenc}
\usepackage{ae,aecompl}

\usepackage{multirow}
\usepackage{breakurl}

\usepackage{graphicx}	
\usepackage{amsmath}	
\usepackage{amssymb}	
\usepackage{array}

\newcommand{\Gaia}{\emph{Gaia}-DR2}
\newcommand{\hii}{H\,\textsc{ii}}
\newcommand{\Bailer}{BJ15}
\newcommand{\Luri}{L18}

\title[{\Gaia} distance to the W3\,Complex]{{\Gaia} distance to the W3\,Complex in the Perseus\,Arm}

\author[F. Navarete et al.]{
Felipe Navarete$^{1}$\thanks{E-mail: navarete@usp.br (FN)},
Phillip~A.~B.~Galli$^{2}$ and
Augusto Damineli$^{1}$
\\
$^{1}$Universidade de S\~ao Paulo, Instituto de Astronomia, Geof\'isica e Ci\^encias Atmosf\'ericas, Rua do Mat\~ao, 1226, 05508-090, Brazil. \\
$^{2}$Laboratoire d'astrophysique de Bordeaux, Univ. Bordeaux, CNRS, B18N, all\'ee Geoffroy Saint-Hillaire, 33615 Pessac, France.
}

\date{Accepted XXX. Received YYY; in original form ZZZ}

\pubyear{2019}

\begin{document}
\label{firstpage}
\pagerange{\pageref{firstpage}--\pageref{lastpage}}
\maketitle

\begin{abstract}
The Perseus Arm is the closest Galactic spiral arm from the Sun, offering an excellent opportunity to study in detail its stellar population. However, its distance has been controversial with discrepancies by a factor of two. Kinematic distances are in the range 3.9-4.2\,kpc as compared to 1.9-2.3\,kpc from spectrophotometric and trigonometric parallaxes, reinforcing previous claims that this arm exhibits peculiar velocities.
We used the astrometric information of a sample of 31 OB stars from the star-forming W3\,Complex to identify another 37 W3 members and to derive its distance from their {\Gaia} parallaxes with {improved} accuracy.
The {\Gaia} distance to the W3\,Complex,2.14\,$^{+0.08}_{-0.07}$\,kpc, coincides with the previous stellar distances of $\sim$\,2\,kpc.
The {\Gaia} parallaxes {tentatively} show differential distances {for different parts of the} W3\,Complex: W3\,Main, located to the NE direction, is at 2.30\,$^{+0.19}_{-0.16}$\,kpc, the W3\,Cluster (IC\,1795), in the central region of the complex, is at2.17$^{+0.12}_{-0.11}$\,kpc, and W3(OH) is at 2.00$^{+0.29}_{-0.23}$\,kpc to the SW direction.
The W3\,Cluster is the oldest region, indicating that it triggered the formation of the other two star-forming regions located at the edges of an expanding shell around the cluster.
\end{abstract}




\section{Introduction}

	The Milky Way is a benchmark to understand the structure and evolution of spiral galaxies in details.
	However, the position of the Solar system, embedded into the Galactic disc, makes it difficult to determine the precise structure of the spiral arms.

In the past decades, most of the Galactic spiral structure studies relied on the position and Galactocentric distance of different sources derived from their radial velocity and inferring a model for the Galactic rotation curve (e.g. the standard Galactic rotation curve based on hydrogen emission from {\hii} regions by \citealt{Georgelin76} and \citealt{Brand93}; star-forming complexes by \citealt{Russeil03}, high-mass star-forming regions from \citealt{Reid14}; and red clump giants from \citealt{Lopez14}).
     Recently, \citet{Reid14,Reid16} reported a view of the spiral structure of the Milky Way based on distances derived from trigonometric parallaxes of masers in high-mass star-forming regions. A complementary view of the Galactic spiral structure, based on the molecular content of high-mass star-forming regions, is presented by \citet[][see their Fig.\,6]{Urquhart14}.
    \citet{Hou14} combined the data obtained for a large number of {\hii} regions, giant molecular clouds and 6.7\,GHz methanol masers to derive a comprehensive model for the galactic rotation curve and spiral arms distribution.
    Up-to-date, the above-mentioned works correspond to the most accurate Galactic rotation curve models available in the literature.

   The assumption of a Galactic rotation model (commonly refereed as kinematic method) relies upon several assumptions of the geometry and motion of the Galactic disc. For instance, the orbits are assumed to be circular, and deviations from circular orbits are often induced by dynamical processes in the Galaxy, such as the propagation of winds from high-mass stars \citep{Kudritzki00}, shocks from supernovae explosions \citep{Zhou16}, or fluctuations in the gravitational potential \citep{Junqueira13}.
    The combination of these effects can result in large non-circular velocities, leading to unrealistic farther kinematic distance determinations. 
    The catalogue of Galactic complexes of {\hii} regions from \citet{Moises11} indicates that half of the structures have kinematic distances larger by a factor of two than their corresponding non-kinematic distances, mostly derived from spectrophotometric analysis and trigonometric parallaxes.

    Spectrophotometric distances are often obtained through spectral type classification of stars using photometry and spectroscopy in the optical or near-infrared (NIR) (e.g. \citealt{Hanson96}, \citealt{Moises11}). This method requires a reasonable modelling of the interstellar medium reddening and an accurate calibration for the spectral type of the stars (e.g. \citealt{Hanson05}).
    The reddening and the spectral classification correspond to the major sources of uncertainty of the spectrophotometric technique.
     The scattering on the calibration of the absolute magnitudes of O-type stars ($\Delta K$\,=\,$\pm$\,0.67\,mag) translates into an error of 30\% over the distance of each source \citep{Blum00}.    
     The uncertainty on the reddening arises from the impossibility of evaluating the local interstellar medium (ISM) for each region.
     Since there are different reddening laws available in the literature (e.g. \citealt{Cardelli89},  \citealt{Stead09} and \citealt{Damineli16}), two extremes are often adopted for obtaining a mean reddening correction and its uncertainty.
     
    On the other hand, distances derived from annual trigonometric parallaxes are not dependent on any modelling of stellar or ISM parameters, offering a direct estimate of the distance to the source.
    In the past decades, Very Long Baseline Interferometry (VLBI) observations have been used to measure the parallax of maser sources with unprecedented accuracy (of about 3\%, \citealt{Xu06}). The available angular resolution of VLBI measurements (about 0.01\,milli-arcseconds at 22\,GHz, \citealt{Hachisuka06}) have favoured the study of the Galactic spiral structure based on bright maser sources.
    
    The Perseus arm is the nearest spiral arm from the Sun, located at the Galactic anti-centre direction.
    The complex of {\hii} regions W3 (hereafter, W3\,Complex) is one of the most prominent star-forming regions located in the second Galactic quadrant ($\ell$\,$\sim$130$^\circ$) and associated with the Perseus arm, offering an excellent opportunity to determine the distance to this spiral arm using both kinematic and non-kinematic methods.

The W3\,Complex is located in between the famous Heart and Soul nebulae, in the Cassiopeia constellation. The nebulae are powered by the W5 and W4+W3 radio sources, respectively, and covers a region of about 5.5\,$\times$\,4 degrees in the sky.
    These active star-forming {\hii} regions trace the Perseus spiral arm in the Galactic anti-centre direction.
        The existence of at least three stellar clusters -- IC\,1848 in W5 \citep{Hoag61}, IC\,1805 in W4 \citep{Vasilevskis65}, and IC\,1795 in W3 \citep{Ogura76} -- combined to the relatively low ISM extinction makes this site strategic to study the formation of massive stars because its distance is prone to be determined with high accuracy.
    
    Previous non-kinematic distances to these stellar clusters, especially IC\,1805 (in W4), indicate values in the range of 2.3--2.4\,kpc \citep{Kwon83,Massey95,Sung17}, while the distance to IC\,1848 (W5) ranges from 1.9 to 2.2 kpc \citep{Becker71,Georgelin76,Chauhan11}. Also, the kinematic distance to W5, $d$\,$\sim$\,3\,kpc, is about 1.5 times larger than its stellar distances \citep{Ginsburg11}.

    The distance to the W3\,complex was derived from its stellar content in the optical \citep{Humphreys78} and in the near-infrared \citep{Navarete11}, leading to values around d\,$\sim$\,2.2\,kpc, in agreement with the non-kinematic distances to W4 and W5.
    The distance to W3 was also derived by using high-angular resolution from VLBI observations to measure the trigonometric parallaxes of maser emission sources (e.g. H$_2$O or CH$_3$OH), leading to distances of d\,$\sim$\,2.0\,kpc \citep{Hachisuka06,Xu06}. 
    However, the kinematic distances to W3, ranging from 2.9 to 4.2\,kpc (\citealt{Reid14} and \citealt{Russeil03}, respectively), are systematically larger than those obtained from the non-kinematic methods mentioned above.
    Table\,\ref{tab:distances_w3} summarises the distances reported in the literature for the W3\,Complex.
    When available, the table also indicates the corresponding sub-structure of the W3\,Complex studied in each case.

\setlength{\tabcolsep}{6pt}
\begin{table}
  \caption{Distances to W3 reported in the literature.}
  \label{tab:distances_w3}
  \begin{tabular}{lcccr}
    \hline
Method 						&   Range   & Region &  $d$ (kpc)                 & 	Ref. \\    
    \hline
Spectrophotometric	        &	Optical	&	b	&	2.18                	&	H78	\\
Spectrophotometric	        &	Optical	&no info.&	2.30\,$\pm$\,0.25	    &	R03	\\
Spectrophotometric          &	Infrared& a,b,c	&	2.20$^{+0.80}_{-0.64}$	&	N11	\\
Trigonometric parallax		&	22\,GHz	&	a	&	1.9\,$\pm$\,0.3	        &	I00	\\
Trigonometric Parallax		&	22\,GHz	&	c	&	2.04\,$\pm$\,0.07   	&	H06	\\
Trigonometric parallax 		&	Radio	&	c	&	1.95\,$\pm$\,0.04   	&	X06	\\
Kinematic distance			&	Radio	&	c	&	2.93$^{+0.68}_{-0.63}$	&	R14	\\
Kinematic distance			&	Radio	&	c	&	3.81$^{+0.96}_{-0.69}$	&	W18	\\
Kinematic distance			&	Radio	&no info.&	4.2$^{+0.7}_{-0.6}$	    &	R03	\\
    \hline
\end{tabular} \\
{\textbf{Notes:}
Regions: $a)$ W3\,Main; $b)$ W3\,Cluster; $c)$ W3(OH).
References: H78 -- \citet{Humphreys78}; I00 -- \citealt{Imai00}; R03 -- \citealt{Russeil03}; H06 -- \citealt{Hachisuka06}; X06 -- \citealt{Xu06}; N11 -- \citealt{Navarete11}; R14 -- \citealt{Reid14}; W18 -- \citealt{Wenger18}.}
\end{table}
\setlength{\tabcolsep}{6pt}

	Most studies of the W3\,Complex have focused on its high-density layer (HDL) \citep[e.g.][]{Feigelson08,Navarete11,Bik12}, which is believed to be the result of the expansion of the {\hii} region driven by the cluster IC\,1805 (i.e. the Heart Nebula), located in W4, into the W3 Giant Molecular Cloud (GMC) \citep{Lada78}.
    The HDL of W3 comprises a variety of {\hii} regions at different evolutionary stages, and a total of 105 OB-type stars were spectroscopically confirmed as W3 members \citep{Kiminki15}.
    The W3\,Complex corresponds to the central part of the HDL, and is divided into three main sub-structures: the central IC\,1795 cluster, W3\,Main located to the NW direction, and W3(OH) to the SE direction.

    An O6.5\,V star (BD+61 411) lies at the centre of the W3\,Complex, corresponding to the main source of the IC\,1795 cluster (hereafter, W3\,Cluster). 
    The W3\,Cluster is the most prominent structure of the W3\,Complex at optical wavelengths with a population of about 2,000 stellar objects \citep{Roccatagliata11}, and is one of the oldest clusters \citep[3-5\,Myr,]{Oey05} within W3. The cluster is located at the central position of an expanding super-bubble with a radius of {$\sim$\,9\arcmin}, suggesting that the feedback from its star-forming activity triggered the subsequent star-forming events on the other sub-structures, such as W3\,Main and W3(OH) (see Fig.\,\ref{fig:w3_map}). 
    
    W3\,Main is located to the NW direction of the W3\,Cluster and is an extended star-forming region, exhibiting a large number of infrared sources and {\hii} regions at different evolutionary stages. 
    \citet{Bik12} investigated the OB stellar population of W3\,Main, identifying 15 OB stars with spectral types between O5\,V and B4\,V. Those authors derived an age spread of 2-3\,Myr for the W3\,Main region, where the most massive star (IRS\,2) is already evolved while other high-mass YSOs (e.g. IRS\,N1) are still deeply embedded in ultra-compact {\hii} regions. Moreover, \citet{Bik12} reported that only in the hyper-compact {\hii} region IRS\,5 the early-type stars are still surrounded by circumstellar material.
   X-ray observations of the pre-main sequence population of W3\,Main \citep{Feigelson08} indicate that the whole structure has a spherical distribution with an angular diameter of $\sim$\,12\arcmin, about twice the size inferred from near-infrared maps.
   
W3(OH) is a slowly expanding shell-like {\hii} region with velocities of 3-5\,km\,s$^{-1}$ \citep{Kawamura98}, located to the SE direction of the W3\,Complex. This region is well-known regarding its stellar population \citep{Bik12} and maser emission \citep{Menten88,Ojha04,Hachisuka06,Xu06}. The presence of early B-type stars at the Main Sequence (\citealt{Navarete11} and \citealt{Kiminki15}) indicates that W3(OH) had a relatively recent star formation episode. 

    Thanks to the second release of the \textit{Gaia} mission \citep[hereafter, {\Gaia},][]{Clementini16,Brown18}, accuracy {values closer to those obtained with} VLBI measurements in the radio domain is now extended for targets visible at optical wavelengths.
    Indeed, the parallax errors for individual sources in the W3\,Complex ranges from 0.02 to 0.16\,milli-arcseconds (mas).
    In this study, we used the {\Gaia} measurements of the W3\,Complex OB stellar population to infer a more precise distance to that region and to disentangle the distances to the sub-regions within the W3\,Complex.

This manuscript is organised as follows.
In Sect.\,\ref{sec_data}, we present the sample of known OB stars in the literature associated with the W3\,Complex and new candidate members selected in our analysis.
In Sect.\,\ref{sec_results}, we analyse the {\Gaia} parallaxes and we determine the distance to the W3\,Complex, together with the evaluation of the distance to each sub-region of the complex.
In Sect.\,\ref{sec_discussion}, we compare our results with previous studies.
Our conclusions are summarised in Sect.\,\ref{sec_conclusions}.

\section{Data}
\label{sec_data}

\subsection{Known members of the OB stellar population of W3}
\label{sec_obstars}

   A total of 48 OB stars within the W3\,Complex, centred at RA\,=\,02:26:32.4, Decl.\,=\,+62:00:27 and covering a {$\sim$\,20\arcmin$\times$20\arcmin} region, were investigated and classified by \citet{Navarete11}, \citet{Bik12} and \citet{Kiminki15} and their properties are listed in Table\,\ref{tab:ob_stars_full}.
  
    The majority of sources (the numbers are given in parenthesis) are found within the three main sub-structures of the W3\,Complex: W3\,Main (10), W3\,Cluster (8), and W3(OH) (4).
    The last 9 OB stars were classified as field stars associated with the HDL region. Although these objects are confirmed members of the W3 complex, they are not likely associated with any of the former sub-structures.
    
\input{table_ob_stars.tex}

\begin{figure*}
	\includegraphics[width=0.8\linewidth]{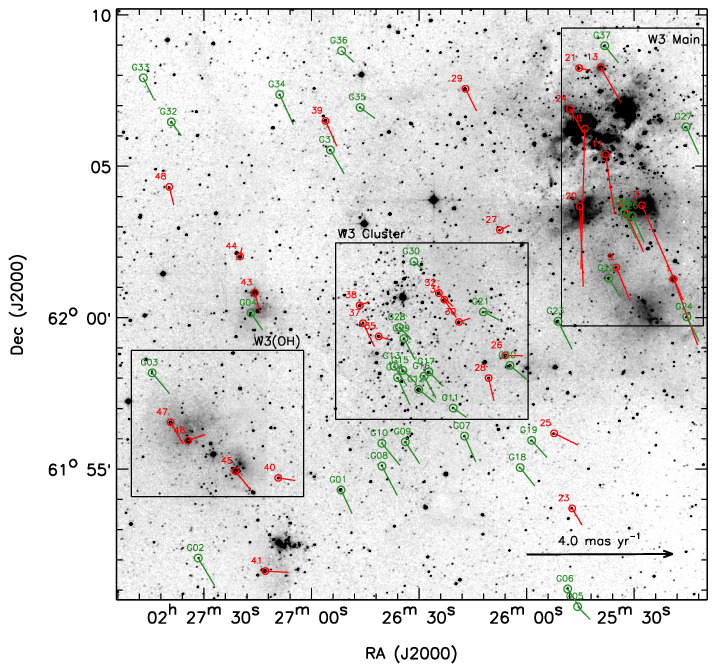} \\[-4.0ex]
    \caption{$K$-band map of the W3\,complex (centred at RA\,=\,02:26:32.4, Decl.\,=\,+62:00:27) overlaid by the OB stellar population with astrometry available in the {\Gaia} catalogue.
    The red circles indicate the position of the OB stars (labelled using the IDs from Table\,\ref{tab:ob_stars_full}), and the arrows indicate the direction and magnitude of their proper motions (a scale is indicated by the black arrow at the bottom right corner of the map).
    The green circles indicate the position of {\Gaia} objects with similar astrometric parameters as the OB stars and candidate members of the W3\,Complex (see Sect.\,\ref{sec_largesample}).
	The black rectangles indicate the location of the W3\,Main, W3\,Cluster and W3(OH). 
    North is up, and East is to the left.
}
    \label{fig:w3_map}
\end{figure*}

	The cross-match between the sources listed in Table\,\ref{tab:ob_stars_full} with the {\Gaia} catalogue can lead to doubtful associations since the positions of both lists are not given for the same epoch.
    Fortunately, the {\Gaia} provides the cross-match between its sources with external large dense surveys, including the \emph{Two Micron All-Sky Survey} \citep[2MASS,][]{Skrutskie06}.
    Thus, we used the 2MASS IDs listed in Table\,\ref{tab:ob_stars_full} to search for the corresponding {\Gaia} counterparts in the \texttt{TMASS\_BEST\_NEIGHBOUR} table of the {\Gaia} archive. Then, we downloaded the {\Gaia} data from the main catalogue.  
     From the initial list of 48 OB stars, only 36 sources have complete information in the {\Gaia} catalogue (i.e. position, proper motion and parallax) required for the present study. 
     The resulting list is presented in Table\,\ref{tab:gaia_parallaxes} and the {\Gaia} sources are indicated in Fig.\,\ref{fig:w3_map}, together with the vectors indicating the direction and magnitude of their proper motions.

\begin{figure}
	\includegraphics[width=\columnwidth]{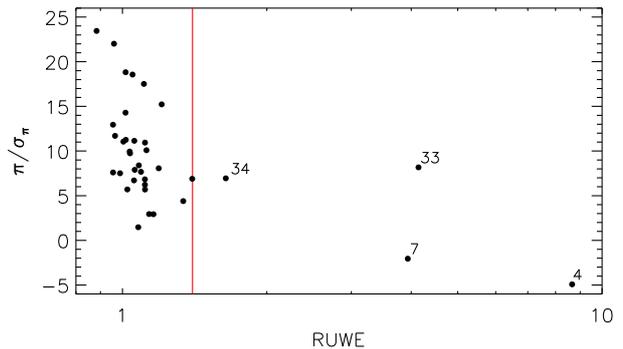} \\[-6.0ex]
    \caption{Parallax over the error ($\pi/\sigma_\pi$) versus the Re-normalised Unit Weight Error (RUWE) of the {\Gaia} sources.
    The vertical red line is placed at RUWE\,=\,1.4, and the labelled sources were excluded from the analysis due to their large RUWE values.}
    \label{fig:w3_separation}
\end{figure}

    We followed the recommendation of \citet{Lindegren18}, we checked the goodness of the astrometric solution of each source by evaluating the re-normalised unit weight error parameter (RUWE), and keeping only those sources with RUWE\,$\leq$\,1.4.
    Figure\,\ref{fig:w3_separation} presents the parallax over error ratio ($\pi/\sigma_\pi$) as a function of the RUWE parameter for all the {\Gaia} objects in the sample. The distribution of the points indicates that the adopted criterion was sufficient to separate sources with good astrometric solutions from those associated with either negative parallaxes or relatively large uncertainties (\#4, \#7, \#33 and \#34, see Fig.\,\ref{fig:w3_separation}).
    In addition, the source \#22 was also excluded from the sample due to the large uncertainty on its parallax ($\pi$\,=\,0.97\,$\pm$\,0.66\,mas).
    These procedures led to a final sample of 31 OB stars with reliable {\Gaia} parallaxes measurements.

\input{table_gaia.tex}

    Figure\,\ref{fig:w3_map} exhibits the large-scale $K_{\rm s}$-band map (at 2.16\,$\mu$m) of the central {20\arcmin$\times$20\arcmin} region of the W3\,Complex, extracted from the 2MASS image archive\footnote{\url{https://irsa.ipac.caltech.edu/applications/2MASS/IM/}}.
    The {\Gaia} sources are overlaid on the map, together with the vectors indicating the direction and magnitude of their proper motions.

\begin{figure}
  \includegraphics[width=\linewidth]{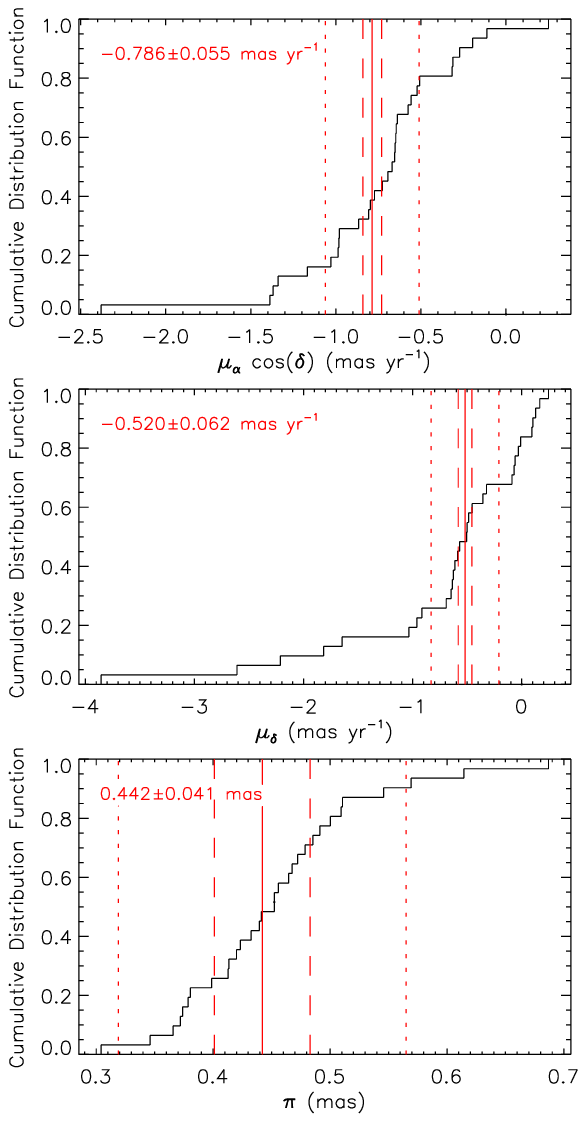} \\[-7.0ex] 
\caption{Cumulative distribution function of the astrometric parameters of the OB stars. The distribution of the proper motions in the right ascension and declination axis are presented in the top and middle panels, respectively, and the parallaxes in the bottom. The mean and its error from Table\,\ref{tab:parallaxes_w3} are presented in the top left corner of each panel and are indicated by the solid and dashed vertical red lines, respectively. The threshold used for selecting the larger sample of {\Gaia} objects are indicated by the dotted red lines (corresponding to 5- and 3-$\sigma$ for the proper motion and parallaxes, respectively).}
    \label{fig:cdf_astrometry}
\end{figure}

Figure\,\ref{fig:cdf_astrometry} presents the cumulative distribution of the proper motions in the right ascension and declination axis, and the parallaxes of the OB stars. The distributions of the proper motions are shifted towards negative values, indicating that most of the {\Gaia} sources are moving in the NW direction. The parallaxes are located within 0.3 and 0.7\,mas, and the naive inversion of their individual values leads to distances in the range of 1.4--3.3\,kpc.

We evaluated the weighted mean mean astrometric parameters of W3 and its sub-regions as:
\begin{equation}
    \langle p \rangle = \sum_{i=1}^{n} w_i p_i
    \label{eq_meanpi}
\end{equation}
{\noindent where the weight, $w_i$, is defined in terms of the uncertainty of the parameter $p$:}
\begin{equation}
    w_i = \frac{ \sigma_{\rm{p},i}^{-2} }{ \sum_{i=1}^{n} \sigma_{\rm{p},i}^{-2} }
    \label{eq_weightpi}
\end{equation}

The error of the weighted mean was obtained by considering the uncertainty on each measurement of the parameter and the spatial correlation between the position of the stars:
\begin{equation}
    \sigma_{\langle \pi \rangle}^2 = \sum_{i=1}^{n} \sum_{j=1}^{n} \left( w_i \sigma_{\rm{\pi},i} \right) \cdot \left( w_j \sigma_{\rm{\pi},j} \right) \cdot V(\theta_{ij})
    \label{eq_errpi}
\end{equation}
\noindent {where $V(\theta_{ij})$ is the spatial covariance function between sources $i$ and $j$, defined as:}
\begin{equation}
  V(\theta_{ij}) = A \cdot \exp \left( -\frac{ \theta_{ij} }{ \Delta\ell } \right)
  \label{eq_spatcorr}
\end{equation}
\noindent with $\theta_{ij}$ being the angular separation between sources $i$ and $j$, $A$\,=\,1 (for $i$\,=$j$, $V(\theta)$\,=\,1), and $\Delta\ell$\,=\,0.5$^\circ$ is defined by \citet{LL:LL-124}.
The weighted mean proper motion of the OB stellar population is $-$0.786\,$\pm$\,0.055 and $-$0.520\,$\pm$\,0.062\,mas\,yr$^{-1}$ and the mean parallax is $\langle \pi \rangle$\,=\,0.442\,$\pm$\,0.041\,mas (with no zero-point correction, see Sect.\,\ref{sec_zp}). These values are indicated in Fig.\,\ref{fig:cdf_astrometry} as the solid and dashed lines.

\subsection{New members of the W3\,Complex based on {\Gaia} data}
\label{sec_largesample}

We selected a larger sample of candidate members of the W3\,Complex based on the astrometric parameters of the well-known OB stars present in the same region.
At first, we have selected all the 2249 {\Gaia} sources located within the FOV presented in Fig.\,\ref{fig:w3_map} which also satisfied the RUWE\,$\leq$\,1.4 criterion (see Sect.\,\ref{sec_obstars}). Then, we narrowed the sample to 99 sources by selecting only those objects with proper motions within the range between $\pm5\sigma$ from the weighted mean proper motion values of the OB stars, that is, $-$1.06\,$\leq$\,$\mu_\alpha\cos{(\delta)}$\,$\leq$\,$-$0.51\,mas\,yr$^{-1}$ and $-$0.83\,$\leq$\,$\mu_\delta$\,$\leq$\,$-$0.21\,mas\,yr$^{-1}$.

These sources were associated with parallaxes with a larger distribution range (from $-$1.6 to 1.3\,mas) when compared to the parallaxes of the OB stars (0.30 to 0.69\,mas). Thus, selected only those objects with parallax measurements within a 3-$\sigma$ limit from the weighted mean parallax value (i.e. 0.319\,$\leq$\,$\pi$\,$\leq$\,0.565\,mas), leading to a sample of 45 objects. This selection contains eight of the sources listed in Table\,\ref{tab:gaia_parallaxes} and 37 new objects. Table\,\ref{tab:gaia_parallaxes_largersample} lists the properties of the 37 additional {\Gaia} objects.
The association between the 31 confirmed members of the W3\,complex and the 37 new candidates selected based on the criteria defined above lead to a final sample of 68 {\Gaia} sources.

\input{table_gaia_newsample.tex}

    We performed a two-sample Kolmogorov-Smirnov test to check test whether the two distributions are different (i.e. the OB star population -- see Sect\,\ref{sec_data} -- and the {\Gaia} selected sources -- see Sect.\,\ref{sec_largesample}).
    The samples are considered statistically different if their KS rank factor is close to 1 and associated with a probability value, $p$ smaller than 0.05 ($p$-values of 0.05, 0.002 and <\,0.001 represent the $\sim$2-, 3- and >\,3-$\sigma$ confidence levels).
    The results are summarised in Table\,\ref{tab:ks_w3}.
    We found that the KS rank factors are about $\sim$0.3 for the proper motions, indicating a relatively small difference between the distribution of the OB stars and the sources listed in Table\,\ref{tab:gaia_parallaxes_largersample}.
    The KS rank factor for the parallax is even smaller, KS\,=\,0.12 and associated with a high $p$-value ($p$\,=\,0.95), indicating a non-significant difference between each distribution. These results support that both the known OB stellar population of W3 and the new {\Gaia} candidate members of W3 are similar in terms of their astrometric information.

\setlength{\tabcolsep}{12pt}
\begin{table}
  \caption{Kolmogorov-Smirnov statistics between the astrometric parameters of the OB stellar population and the {\Gaia} candidate members of W3.}
  \label{tab:ks_w3}
  \begin{tabular}{lcccc}
    \hline 
Parameter&	$\pi$                       &$\mu_\alpha\cos(\delta)$   &$\mu_\delta$            	&$\mu$                   	\\
\hline
KS	    &		0.12   &   	0.33   &   0.32   &	0.32      \\
$p$     &       0.95   &   0.04       &   0.05   &   0.05   \\
\hline
\end{tabular} \\
{\textbf{Notes:} The KS test and its probability ($p$) are shown for each parameter. KS values closer to 1 and $p$-values smaller than 0.05 indicate that the distributions are significantly different.}
\end{table}
\setlength{\tabcolsep}{6pt}

Finally, we associated the new sample of 68 stars with the sub-regions of the W3\,Complex by using the same position criteria indicated in Fig.\,\ref{fig:w3_map}. We found that 16 sources are within W3\,Main, 20 are associated with W3\,Cluster, five are located in W3(OH), and 27 are likely not associated with any of the former sub-structures (field stars).

\section{Results}
\label{sec_results}

\subsection{Systematic effects on {\Gaia}-parallaxes}
\label{sec_systematics}

According to \citet{Lindegren18}, the systematic errors in the {\Gaia} parallaxes are estimated to be less than 0.1\,mas.
The published uncertainties of the {\Gaia} parallaxes correspond to internal errors only and do not consider systematic or external errors \citep{Lindegren18,LL:LL-124}. 
By following the recommendation of \citet{LL:LL-124}, the total uncertainty on the parallax, $\sigma_\pi$, is a combination of the internal parallax error ($\sigma_{\rm i}$) and an external error ($\sigma_{\rm e}$), given by:
\begin{equation}
    \sigma_\pi = \sqrt{ k^2 \sigma_{\rm i}^2 +  \sigma_{\rm e}^2 }
    \label{eq_pi_error}
\end{equation}
\noindent where $k$\,=\,1.08, and $\sigma_{\rm e}$ depends on the brightness of the source ($\sigma_{\rm e}$\,=\,0.021\,mas for $G$\,$\leq$\,13\,mag, and $\sigma_{\rm e}$\,=\,0.043\,mas for $G$\,>\,13\,mag). Similarly, the total error of the proper motions and position were also evaluated using Eq.\,\eqref{eq_pi_error}, with $\sigma_{\rm e}$\,=\,0.033\,mas\,yr$^{-1}$ or 0.066\,mas\,yr$^{-1}$ (and in case of the position, 0.016\,mas or 0.033\,mas) for $G$\,$\leq$\,13\,mag or $G$\,>\,13\,mag, respectively.

Another systematic effect in the {\Gaia} parallaxes corresponds to the existence of a global zero-point (ZP) correction of their values, which should be applied to all parallaxes before any further interpretation of their results.
    For the analysis of the W3 parallaxes, we adopted the ZP correction derived by \citet{Graczyk19}, $\pi_{\rm ZP}$\,=\,$-$0.031\,$\pm$\,0.011\,mas, based on the analysis of an all-sky sample of 81 galactic eclipsing binary stars.
    The effects of different ZP offsets on the {\Gaia} parallaxes are further discussed in Sect.\,\ref{sec_zp}).      
     
\subsection{Weighted mean astrometric parameters}
\label{sec_mean_parameters}

    After correcting for the global zero-point, the parallaxes of the stars ranged from 0.335 to 0.718\,mas.
Using Eqs.\,\eqref{eq_meanpi} and \eqref{eq_errpi}, we evaluated the weighted mean parallax of the W3\,complex and its error as $\langle \pi \rangle$\,=\,0.473\,$\pm$\,0.041\,mas considering only the OB stars, and $\langle \pi \rangle$\,=\,0.477\,$\pm$\,0.044\,mas when considering the larger sample of 68 stars. Similarly, we obtained the weighted mean position and proper motions of the W3 and its sub-regions.
These values are reported in Table\,\ref{tab:parallaxes_w3}, together with the corresponding distances using the naive inversion of the parallaxes. 
    However, the evaluation of distances by inverting parallaxes is only valid in the absence of uncertainties. That is not the case for {\Gaia} parallaxes since they have intrinsic uncertainties provided in the {\Gaia} catalogue and are also affected by systematic errors as previously mentioned in Sect.\,\ref{sec_systematics}.
    For this reason, we adopted a Bayesian inference method to derive the distance to the W3\,complex and its sub-structures. The description of this method and the results are presented as follows.

\setlength{\tabcolsep}{9pt}
\begin{table*}
  \caption{Mean parameters and distance to the W3\,Complex and its main sub-regions based on the OB stars and the larger {\Gaia} sample.}
  \label{tab:parallaxes_w3}
  \begin{tabular}{lc|cccc|ccc}
  \hline
  \hline
\multicolumn{9}{c}{OB stars} \\
    \hline 
Region		&	N	&	$\langle\alpha\rangle$  &	$\langle\delta\rangle$  &	$\langle\mu_\alpha\cos(\delta)\rangle$  &	$\langle\mu_\delta\rangle$	&	$\langle\pi\rangle$&	${\langle d \rangle}$ 	&	d$_{\rm EDSD}$ 	 \\
    		&	    &	(deg)  &	(deg)  &	(mas\,yr$^{-1}$)  &	(mas\,yr$^{-1}$)    &	(mas)   &   (kpc)  &   (kpc) 	\\
\hline
W3	        &	31	&	36.598\,$\pm$\,0.029 &   62.011\,$\pm$\,0.030    &   -0.786\,$\pm$\,0.055    &   -0.520\,$\pm$\,0.062    &   0.473\,$\pm$\,0.041 &   2.11$^{+0.20}_{-0.17}$   &   2.23$^{+0.15}_{-0.14}$	\\
\hline
W3\,Main 	&	10 	&	36.398\,$\pm$\,0.036 	&	62.076\,$\pm$\,0.036 	&	-1.000\,$\pm$\,0.066 	&	-1.312\,$\pm$\,0.073 	&	0.444\,$\pm$\,0.064 	&	2.25$^{+0.38}_{-0.28}$	&	2.46$^{+0.25}_{-0.21}$	\\
W3\,Cluster &	8 	&	36.615\,$\pm$\,0.030 	&	61.995\,$\pm$\,0.030 	&	-0.661\,$\pm$\,0.056 	&	-0.179\,$\pm$\,0.063 	&	0.460\,$\pm$\,0.055 	&	2.17$^{+0.29}_{-0.23}$	&	2.07$^{+0.18}_{-0.15}$	\\
W3(OH) 		&	4 	&	36.852\,$\pm$\,0.045 	&	61.924\,$\pm$\,0.046 	&	-0.889\,$\pm$\,0.079 	&	-0.28\,$\pm$\,0.10 	&	0.534\,$\pm$\,0.078 	&	1.87$^{+0.32}_{-0.24}$	&	2.00$^{+0.30}_{-0.23}$	\\
Field 		&	9 	&	36.674\,$\pm$\,0.030 	&	61.996\,$\pm$\,0.030 	&	-0.706\,$\pm$\,0.056 	&	-0.249\,$\pm$\,0.064 	&	0.498\,$\pm$\,0.055 	&	2.01$^{+0.25}_{-0.20}$	&	2.19$^{+0.23}_{-0.19}$	\\
    \hline
    \hline
\multicolumn{9}{c}{Larger {\Gaia} sample} \\
    \hline
Region		&	N	&	$\langle\alpha\rangle$  &	$\langle\delta\rangle$  &	$\langle\mu_\alpha\cos(\delta)\rangle$  &	$\langle\mu_\delta\rangle$	&	$\langle\pi\rangle$&	${\langle d \rangle}$ 	&	d$_{\rm EDSD}$ 	 \\
    		&	    &	(deg)  &	(deg)  &	(mas\,yr$^{-1}$)  &	(mas\,yr$^{-1}$)    &	(mas)   &   (kpc)  &   (kpc) 	\\
\hline
W3	    &	68	&	36.606\,$\pm$\,0.032	&	62.005\,$\pm$\,0.032	&	-0.767\,$\pm$\,0.059	&	-0.528\,$\pm$\,0.068	&	0.477\,$\pm$\,0.044	&	2.10$^{+0.21}_{-0.18}$	&	 2.14$^{+0.08}_{-0.07}$   \\
\hline
W3 Main	&	16	&	36.382\,$\pm$\,0.033	&	62.074\,$\pm$\,0.033	&	-0.848\,$\pm$\,0.061	&	-1.327\,$\pm$\,0.068	&	0.458\,$\pm$\,0.045	&	2.18$^{+0.24}_{-0.20}$	&	 2.30$^{+0.19}_{-0.16}$   \\
W3 Cluster&	20	&	36.629\,$\pm$\,0.029	&	61.996\,$\pm$\,0.030	&	-0.696\,$\pm$\,0.056	&	-0.132\,$\pm$\,0.063	&	0.471\,$\pm$\,0.040	&	2.12$^{+0.20}_{-0.17}$	&	 2.17$^{+0.12}_{-0.11}$   \\
W3(OH)	&	5	&	36.826\,$\pm$\,0.035	&	61.921\,$\pm$\,0.035	&	-0.779\,$\pm$\,0.064	&	-0.226\,$\pm$\,0.075	&	0.506\,$\pm$\,0.048	&	1.98$^{+0.21}_{-0.17}$	&	 2.00$^{+0.29}_{-0.23}$   \\
W3 Field&	27	&	36.717\,$\pm$\,0.030	&	61.997\,$\pm$\,0.030	&	-0.762\,$\pm$\,0.055	&	-0.799\,$\pm$\,0.064	& 0.504\,$\pm$\,0.042	&	1.98$^{+0.18}_{-0.15}$	&	 2.08$^{+0.11}_{-0.10}$   \\  
\hline
\end{tabular} \\
{\textbf{Notes:}
The columns are as follows:
(1) sub-region of the W3\,Complex;
(2) number of sources;
(3) mean right ascension and the corresponding error (in degrees);
(4) mean declination and the corresponding error (in degrees);
(5) mean proper motion on the right ascension direction and the corresponding error (in mas\,yr$^{-1}$);
(6) mean proper motion on the declination direction and the corresponding error (in mas\,yr$^{-1}$);
(7) mean parallax and the corresponding error (in mas);
(8) distance inferred from the inversion of the mean parallax (in kpc);
(9) distance (in kpc) inferred from the multiple-source Bayesian Inference method (see Sect.\,\ref{sec_gaia_w3}).}
\end{table*}
\setlength{\tabcolsep}{6pt}

\subsection{Distance to the W3\,Complex}
\label{sec_gaia_w3}

   The distance to the W3\,Complex and its sub-structures were calculated by following the tutorials provided in the {\Gaia} archive\footnote{\url{https://gea.esac.esa.int/archive/}} (see, e.g. \citealt{Luri18}, hereafter \Luri).
   We adopted the Exponentially Decreasing Space Density (EDSD) prior from \citet[][hereafter, \Bailer]{Bailer-Jones15}, given by:
\begin{equation}
    P(r) = \frac{r^2}{2L^3} \exp\left(  -\frac{r}{L} \right) 
\end{equation}
\noindent where $r$ is the distance and $L$ is the length scale of the distribution. We adopted the standard value $L$\,=\,1.35\,kpc from \citet{Astraatmadja16}, which was also used for estimating distances from {\Gaia}-DR1 parallaxes.
    We note that the EDSD prior is appropriate for the general, old Galactic population, which has a different spatial Galactic distribution when compared to the young stellar population of the Galactic disc.
For this reason, we tested our results using a self-gravitating, isothermal Galactic disc prior by \citet{Apellaniz01}, based on the analysis of Hipparcos data of the Galactic OB stellar population . The tests indicate that the usage of one prior or another has not substantially changed our results (for details, see Appendix\,\ref{appendix_priors}).
To account for the correlation between each member of the W3\,complex, we used the \textit{multiple-source Bayesian Inference} procedure from the {\Gaia} archive, which assumes that the position ($\alpha$, $\delta$), the parallax ($\pi$) and its error ($\sigma_\pi$) are correlated quantities.

\begin{figure}
  \includegraphics[width=\linewidth]{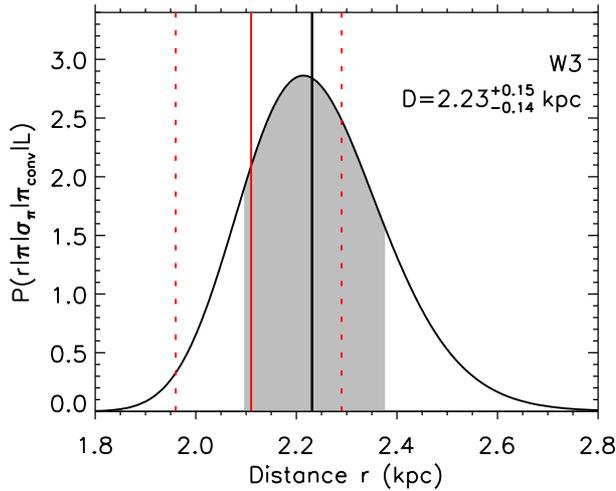}
 \\[-5.0ex] 
\caption{Probability density function (PDF) of the {distance to the W3\,Complex, based on the sample of OB stars.} The median value of the PDF is indicated by the black vertical line. The errors correspond to the 1-$\sigma$ (68\%) confidence interval (shaded grey area). {For comparison, the solid and dashed red lines indicate the distance and its error obtained from the naive inversion of the mean parallax reported in Table\,\ref{tab:parallaxes_w3}.}
}
    \label{fig:pdf}
\end{figure}

Figure\,\ref{fig:pdf} exhibits the probability density function (PDF) of the distance to the W3\,complex {using the EDSD prior} and considering all the {31} OB stars selected as described in Sect.\,\ref{sec_obstars}.
    We note that the PDF exhibits a slight asymmetric profile, elongated towards larger distance values.
    For this reason, {\Bailer} and {\Luri} adopted the median and a 90\% confidence interval (5\% and 95\%) for evaluating the distance and its errors. However, the errors on the previous distance estimates for W3 are based on a 1-$\sigma$ interval (roughly corresponding to a confidence interval of 68\%). Therefore, we adopted a narrower, 68\% confidence interval to better compare our results with those available in the literature (see Table\,\ref{tab:distances_w3}).

    The median value and the 1-$\sigma$ (68\%) confidence interval of the PDF shown in Fig.\,\ref{fig:pdf} correspond to the distance of {2.23$^{+0.15}_{-0.14}$\,kpc}.
    {This value} is relatively larger than the distance derived from the naive inversion of the mean parallax of the OB stars, {d\,=\,2.11$^{+0.20}_{-0.17}$\,kpc}, shown as the solid red line in Fig.\,\ref{fig:pdf}.
    We note that the {accuracy} of the distance evaluated using the {\Gaia} parallaxes and the Bayesian inference method is around 7\% at 1-$\sigma$, {closer to the accuracy of the distances derived from radio parallax measurements \citep[about 3\%, e.g. ][]{Hachisuka06,Xu06} than those obtained through, e.g., the spectrophotometric method (less accurate than 10\%, see Table\,\ref{tab:distances_w3}.)}

\begin{figure}
  \includegraphics[width=\linewidth]{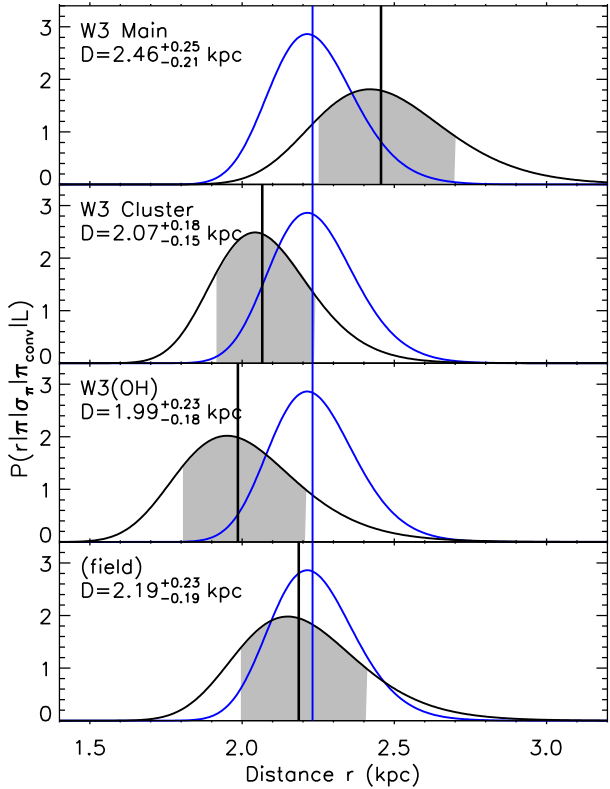} \\[-6.0ex] 
    \caption{PDF of the distance {to each sub-region of the W3\,Complex, based on the sample of OB stars. From top to bottom:} W3 Main, W3\,Cluster, W3(OH), and sources located out of the regions above.
    The distance and its errors are derived from the median of the PDF (vertical line) and the 1-$\sigma$ (68\%) confidence interval (shaded area).
    For comparison, the PDF considering all the OB stars and its median distance are presented as the blue curve and the vertical blue line, respectively.
}
    \label{fig:pdf_regions}
\end{figure}

  Figure\,\ref{fig:pdf_regions} presents the PDF of {the distance} to each sub-structure of the W3 complex, {obtained using the EDSD prior and considering only the OB stars within the corresponding {black} box presented in Fig.\,\ref{fig:w3_map}. The distances derived from the PDFs are listed in Table\,\ref{tab:parallaxes_w3}. }
    In the top panel, we present the PDF of the distance to W3\,Main, located in the NW region of the W3\,Complex (see Fig.\,\ref{fig:w3_map}).
    The PDF exhibits a broad profile, extending from 2.0 to $\sim$\,3.2\,kpc.
    {The median and the 1-$\sigma$ confidence interval of the PDF leads to the distance of $d$\,=\,2.46$^{+0.25}_{-0.21}$\,kpc.}
    The distance to W3\,Main is relatively larger than the distance inferred for the whole W3\,Complex {(overlaid as the blue curve)}, suggesting that this sub-region is likely located at the outer edge of the W3\,Complex.
    The PDF of the distance to the W3\,Cluster is offset towards smaller distances and is relatively narrower when compared to the observed for the W3\,Main. {The median and the 1-$\sigma$ confidence interval of the PDF leads to the distance of {d\,=\,2.07$^{+0.18}_{-0.15}$\,kpc} to the W3\,Cluster.}
    The distribution probability of the distance to W3(OH) exhibits the broadest profile shown in Fig.\,\ref{fig:pdf_regions}, ranging from $\sim$\,1.4 to $\sim$\,2.9\,kpc. 
    The distance to W3(OH), {d\,=\,2.00$^{+0.30}_{-0.23}$\,kpc}, places this sub-region at the nearest side of the W3\,complex.
    Compared to the other sub-regions, the larger errors on the distance of W3(OH) may be caused by the relatively small sample of stars (only four) with measured parallaxes.
    Finally, the PDF {of the distance to the} OB stars classified as field stars is shown in the bottom panel of Fig.\,\ref{fig:pdf_regions}.
    These stars are confirmed members of the W3\,complex but {are not clearly} associated with the former sub-regions. 
    Despite that, and assuming that these stars are indeed members of W3, their position and parallaxes are still correlated.
    The distance derived from their PDF is {d\,=\,2.19$^{+0.23}_{-0.19}$\,kpc}.

\begin{figure}
  \includegraphics[width=\linewidth]{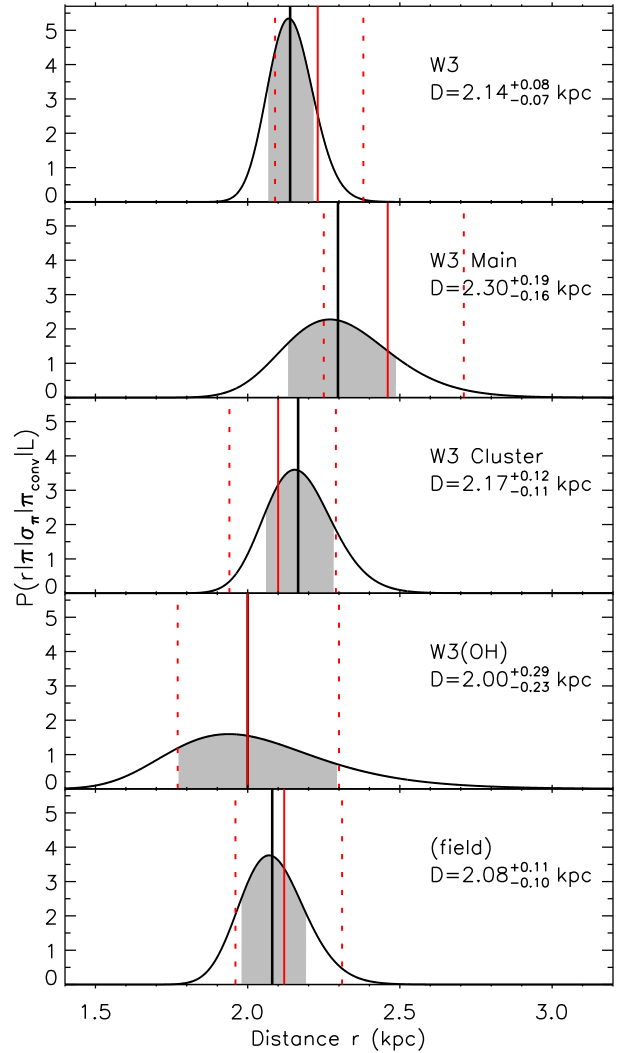} \\[-5.0ex] 
    \caption{PDF of the {distance to the W3\,Complex and its sub-regions based on the larger sample of {\Gaia} sources} located in the following regions (from top to bottom): the whole sample (W3), W3 Main, W3\,Cluster, W3(OH), and sources located out of the regions above (field).
    The distance and its errors are derived from the median of the PDF (vertical line) and the 1-$\sigma$ (68\%) confidence interval (shaded area). In each panel, the median distance and its error derived using the OB stars are indicated by the solid and dashed lines, respectively.
}
    \label{fig:pdf_regions_larger}
\end{figure}

{Figure\,\ref{fig:pdf_regions_larger} presents the PDFs of the distances based on the larger sample of 68 {\Gaia} objects (see Sect.\,\ref{sec_largesample}).
The plots also indicate the distances and their errors obtained using the OB stars and presented in Figs.\,\ref{fig:pdf} and \ref{fig:pdf_regions}.}
    The top panel indicates that the distance to the W3\,Complex is $d$\,=\,2.14$^{+0.08}_{-0.07}$\,kpc when considering the larger sample. we found that the distance using the larger sample is determined with higher accuracy (from $\sim$7 to $\sim$4\%) when compared to the distance obtained with the OB stars, $d$\,=\,2.23\,$^{+0.15}_{-0.14}$\,kpc (see Fig.\,\ref{fig:pdf}).
    
    As previously found in Fig.\,\ref{fig:pdf_regions}, the distances to each sub-region of the W3\,Complex using the large sample of {\Gaia} objects also indicate that they are located at different positions within the high-density layer of W3: W3(OH) is at the inner edge of the W3\,Complex, at 2.00$^{+0.29}_{-0.23}$\,kpc; W3\,Cluster is at the central region at 2.17$^{+0.12}_{-0.11}$\,kpc; and W3\,Main is at the far, outer edge of the W3\,Complex, at $d$\,=\,2.30$^{+0.19}_{-0.16}$\,kpc.

\section{Discussion}
\label{sec_discussion}

Previous works have reported the distance to the W3 Complex based on the individual distances of a few stellar objects \citep[e.g.][]{Humphreys78,Navarete11}, or based on the trigonometric parallax of maser {sources} \citep[e.g.][]{Xu06,Hachisuka06} located in W3(OH).
The trigonometric parallax is the most direct {estimate of the distance to an object} and does not require any assumptions.
On the other hand, {methods such as the spectrophotometric analysis} require a good interpretation of the physical parameters of the {sources} and the conditions of the interstellar medium in the line-of-sight.
    {The {\Gaia} second data release offers the unique opportunity to derive the distance to the W3\,Complex in the optical range with {closer accuracy of} the distances obtained through trigonometric parallax of masers in the GHz regime ($\sim$3\%, e.g. \citealt{Hachisuka06} and \citealt{Xu06}).}
    {Despite of that, we note that the accuracy achieved by {\Gaia} is still not high enough for beating, for example, the quality of VLBI trigonometric parallax measurements. By the time {\it Gaia}-DR3 or DR4 arrives, the accuracy of their measurements might be similarly accurate or even surpass the VLBI measurements.}
    
   The distances derived from the Bayesian inference method using the Exponentially Decreasing Space Distribution prior from {\Bailer} are systematically larger than those obtained from the naive inversion of the weighted mean parallaxes (see Table\,\ref{tab:parallaxes_w3}). {The adoption of other priors, such as the self-gravitating isothermal Galactic disc from \citet{Apellaniz01} leads to similar distances than those derived using the EDSD prior (see Appendix\,\ref{appendix_priors}).}
    As our final results, we adopted the distances derived using the larger sample of {\Gaia} W3 members (Sect.\,\ref{sec_largesample}), {the EDSD prior and} the Bayesian inference method, since it takes the uncertainties of the individual parallax measurements into account, and this prior has been widely used by the {\it Gaia} team.
   {In Sect\,\ref{sec_zp}, we further discuss the influence of the zero-point correction of the {\Gaia} parallaxes on our results.
   The analysis of the {\Gaia} parallaxes indicates that the sub-regions of the W3\,Complex are likely located at different distances in our line-of-sight. These results are discussed in Sect.\,\ref{sec_3d}.
   Finally, we compare the distances to W3 based on the {\Gaia} parallaxes with previous estimates from the literature in Sect.\,\ref{sec_discussion_distance}.}

\subsection{Dependency of the zero-point correction on the distance determination}
\label{sec_zp}

   The major source of systematic errors on the {\Gaia} parallaxes is related to the zero-point correction.
   \citet{Lindegren18} reported that the {\Gaia} parallaxes are systematically smaller than the real values. They derived a zero-point correction of $\pi_{\rm ZP}$\,=\,$-$0.03\,mas, adopted by the {\Gaia} Team as the global ZP offset for the entire {\Gaia} catalogue.
    Despite that, recent works based on the study of different type of Galactic objects have reported larger ZP offsets, between $-$0.08 and $-$0.05\,mas (e.g. \citealt{Riess18}, \citealt{Zinn18} and \citealt{Stassun18}), suggesting that the parallaxes measured by the {\Gaia} are even smaller than predicted by \citet{Lindegren18}.
    
    More recently, \citet{Graczyk19} analysed a sample of 81 eclipsing binaries with measured {\Gaia} parallaxes and obtained a global ZP offset of $\pi_{\rm ZP}$\,=\,$-$0.031\,$\pm$\,0.011\,mas, confirming the value suggested by \citet{Lindegren18}.
    In Sect.\,5 of \citet{Graczyk19}, those authors comment on the discrepancy of the ZP correction adopted by the {\Gaia}\,Team and those obtained by other authors. They claimed that such differences are either affected by systematic effects on the relations used for modelling the stellar parameters that are difficult to assess, or are merely specific for a particular region of the sky (e.g. \citealt{Zinn18} derived a ZP offset based on a sample of objects located in the \textit{Kepler} field at RA\,$\sim$\,19:23:00, Decl\,$\sim$\,+44:30:00).
    Table\,\ref{tab:gaia_zp} summarises the recent {\Gaia} zero-point corrections available in the literature, together with the targets and the region of the sky used for each analysis.

\setlength{\tabcolsep}{3pt}
\begin{table}
  \centering
  \caption{Global zero-point offsets available in the literature for {\Gaia} parallaxes.}
  \label{tab:gaia_zp}
\begin{tabular}{lll}
    \hline
\multicolumn{1}{c}{ZP offset}  &	Sources			&	 Reference \\
\multicolumn{1}{c}{(mas)}        &			    	&	  \\
    \hline
$-$0.030		&	QSOs (all sky)				&	\citet{Lindegren18} \\
$-$0.031(11)	&	Eclipsing Binaries (all sky)&	\citet{Graczyk19}   \\
$-$0.046(14)	&	Cepheids (all sky)			&	\citet{Riess18}     \\
$-$0.050(03)	&	RC stars (Kepler field$^1$)	&	\citet{Zinn18}      \\
$-$0.053(03)	&	RGB stars (Kepler field$^1$)&	\citet{Zinn18}      \\
$-$0.056(11)	&	{\Gaia} RV sample             &	\citet{Schoenrich19}\\
$-$0.073(34)	&	YSOs in Orion$^2$   	    &	\citet{Kounkel18}   \\
$-$0.082(33)	&	Eclipsing Binaries (all sky)&	\citet{Stassun18}   \\
\hline
\end{tabular}
{\textbf{Notes:}
When available, the errors on the last two digits are given within the parenthesis.
The central position of the fields are:
1) RA=19h23m, Decl=+44d30\arcmin;
2) RA=05h40m, Decl=-10d00\arcmin.
}
\end{table}
\setlength{\tabcolsep}{6pt}

We further tested {the influence of} the ZP correction on the distance to the W3\,Complex {by using the parallaxes of all the OB stars listed in Table\,\ref{tab:parallaxes_w3}, and} adopting three extreme cases:
$a)$ no ZP-correction,
$b)$ the ZP correction from \citet[][$-$0.030\,mas]{Lindegren18}, and
$c)$ the extreme case derived by \citet[][$-$0.082(33)\,mas]{Stassun18}.

\begin{figure}
\centering
  \includegraphics[width=\linewidth]{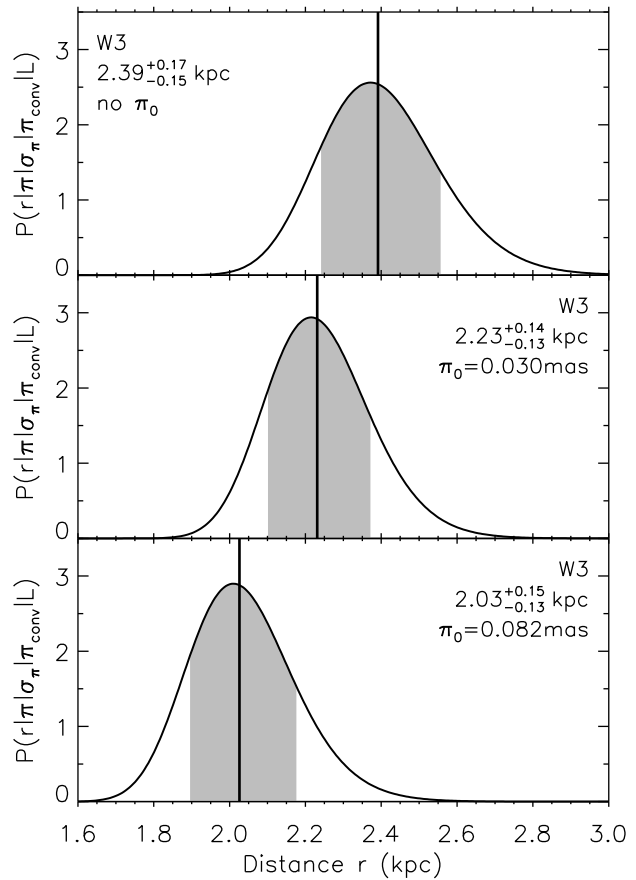} \\[-3.0ex] 
\caption{Probability density function (PDF) of the distance estimate with correlated parallaxes. The median value of the PDF is indicated by the black vertical line. The errors correspond to the 1-$\sigma$ (68\%) confidence interval (shaded grey area). Each PDF was obtained assuming different zero-point correction values: no ZP correction (top), zero-point correction of 0.03\,mas (middle) and ZP correction of 0.082\,$\pm$\,0.033\,mas (bottom).}
    \label{fig:pdf_zp}
\end{figure}

Figure\,\ref{fig:pdf_zp} presents the PDF of the distance to the W3\,Complex after zero-point correcting using the offsets stated above.
{We found that the PDFs are systematically offset towards smaller distances as a function of the ZP value. 
   The median distances and the 1-$\sigma$ uncertainties are:
$(a)$ 2.39$^{+0.17}_{-0.15}$\,kpc,
$(b)$ 2.23$^{+0.14}_{-0.13}$\,kpc, and
 $(c)$ 2.03$^{+0.15}_{-0.13}$\,kpc.
These results indicate that the ZP correction plays a critical role in the correct evaluation of the distances based on {\Gaia} parallaxes, especially for distances at kiloparsec scales, such as the case of W3.}
   
\subsection{The three dimensional structure of the W3 complex and its hierarchical star formation history}
\label{sec_3d}

    The analysis of the OB stellar population of the W3\,Complex and its main sub-regions {tentatively} suggests that the high-density layer of W3 exhibits a complex three-dimensional structure. The W3\,Complex is elongated in the SE-NW direction, where the inner edge is represented by W3(OH) at {$d$\,=\,1.99$_{-0.18}^{+0.23}$\,kpc}, and the outer edge corresponds to the location of W3\,Main, at {$d$\,=\,2.46$_{-0.21}^{+0.25}$\,kpc}. The W3\,Cluster (IC\,1795) is located at the central region of the complex, at {$d$\,=\,2.07$_{-0.15}^{+0.18}$\,kpc}.
    These results are confirmed with the analysis of a larger sample of 68 {\Gaia} sources, which includes the 31 OB stars and 37 new sources selected through astrometric information of the W3\,Complex. When considering the larger sample, the distances to W3(OH), W3\,Cluster and W3\,Main are {$d$\,=\,2.00$_{-0.23}^{+0.29}$\,kpc}, {$d$\,=\,2.17$_{-0.11}^{+0.12}$\,kpc}, and {$d$\,=\,2.30$_{-0.16}^{+0.19}$\,kpc}, respectively.
    
    The projected position on the sky combined with the distance of W3\,Cluster places the oldest structure of the HDL \citep[3-5\,Myr][]{Oey05} at the centre of the W3\,Complex.
    In a larger context, the three dimensional structure of the W3\,Complex is consistent with the hierarchical scenario proposed by \citet{Oey05} and \citet{Roman15}, who suggested that the W3\,Cluster (IC\,1795) triggered the star-forming process and consecutive formation of W3(OH) and W3\,Main{, located at the edges of a shell surrounding the W3\,Cluster. These regions have ages of about 2-3\,Myr \citep{Navarete11, Bik12} and are actively forming stars (cf. \citealt{Bik12} and \citealt{Roman15}).
    In addition, \citet{Oey05} also suggested that the W3\,Cluster formation was induced by an earlier burst of star formation in the W4 region, from which the first event of star formation occurred about 6-10\,Myr ago.}

\subsection{The distance to W3 and the velocity discrepancy in the Perseus Arm}
\label{sec_discussion_distance}

\begin{figure*}
\includegraphics[width=0.75\linewidth]{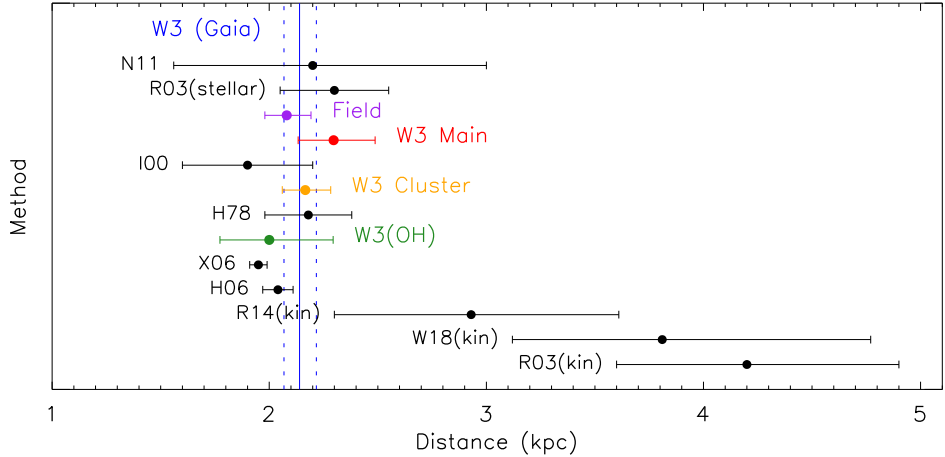} \\[-4.0ex]
\caption{Comparison between distances to the W3\,Complex derived in this work and those reported in the literature.
    The vertical {solid and dashed blue lines} indicate the {\Gaia} distance {to the W3\,Complex and its errors, when considering the large sample of {\Gaia} sources}.
    The {\Gaia} distances to each sub-structure {and their uncertainties} are shown by the coloured dots and their error bars.
    The black {dots} and associated errors correspond to the distances reported in Table\,\ref{tab:distances_w3}, together with the references shown in the plot.
}
    \label{fig:w3_distances} 
\end{figure*}

    Figure\,\ref{fig:w3_distances} compares all the distance determinations for the W3 complex reported in Table\,\ref{tab:distances_w3} with those derived from the {\Gaia} parallaxes using the methodology of {\Bailer} (see last column of Table\,\ref{tab:parallaxes_w3}).
    The distances {from the literature, reported in} Table\,\ref{tab:distances_w3}, are indicated right below the {\Gaia} distance to each W3 sub-structure.
    
    The {\Gaia} distance to the W3 complex, {$d$\,=\,2.05\,$^{+0.10}_{-0.09}$\,kpc} (indicated by the blue solid line), is compatible within 1-$\sigma$ with most of the previous non-kinematic distances. {Excluding the distances derived from the trigonometric parallax of maser sources (e.g. \citealt{Hachisuka06} and \citealt{Xu06}) the error bars of the {\Gaia} distance to the entire W3\,Complex is significantly smaller than the distances obtained from other works.}
    In addition, the most accurate distances based on the annual parallax of maser emission sources in W3(OH) (1.95\,$\pm$\,0.04 and 2.04\,$\pm$\,0.07\,kpc) are in agreement with the {\Gaia} distance to W3(OH), {$d$\,=\,2.07$_{-0.16}^{+0.19}$\,kpc}, derived using the parallaxes of 5 OB stars associated to that region.
    
    The agreement between the distances available in the literature and the relatively more accurate {\Gaia} distance allow us to confirm that the OB-type stars listed in Table\,\ref{tab:ob_stars_full} are indeed high-mass stars located at distances of around $\sim$\,2\,kpc.

    {In a large-scale context, the distance to W3 is also consistent with the distances to the massive clusters located in W4 (IC\,1805) and W5 (IC\,1848), at distances of 2.3-2.4 \citep{Sung17} and 1.9-2.2\,kpc \citep{Chauhan11}, respectively.
    These three complexes cover a considerable portion of the Perseus arm of about 3\,degrees (from 134$^\circ$ to 137$^\circ$) in the Galactic anti-centre direction, allowing us to set a firm location of the Perseus spiral arm with high confidence.}
    
   {At a distance of $\sim$\,2.1\,kpc, the mean proper motion of the W3\,Complex ($\langle\mu_\alpha\cos(\delta)\rangle$\,=\,$-$0.786\,$\pm$\,0.055\,mas\,yr$^{-1}$ and $\langle\mu_\delta\rangle$\,=\,$-$0.520\,$\pm$\,0.062\,mas\,yr$^{-1}$, see Table\,\ref{tab:parallaxes_w3}) leads a to linear velocity of 10.4$\pm$0.9\,km\,s$^{-1}$ roughly oriented in the SW direction.
   We evaluated the mean proper motion of the W3 complex in Galactic coordinates using the transformations available in Sect.\,3.1.7 of the {\Gaia} Documentation release 1.1\footnote{\url{https://gea.esac.esa.int/archive/documentation/GDR2/}}. The proper motions in Galactic coordinates are $\mu_\ell \cos(b)$\,=\,$-$0.527\,mas\,yr$^{-1}$ and $\mu_b$\,=\,$-$0.768\,mas\,yr$^{-1}$.
   The linear velocity components are $v_\ell$\,=\,$-$5.6\,km\,s$^{-1}$ and  $v_b$\,=\,$-$8.2\,km\,s$^{-1}$, with a position angle of $\sim215^\circ$ in relation to the North galactic pole and increasing $\ell$ values.
   Such velocities indicate that the W3\,Complex has a relatively large tangential motion perpendicularly to the Galactic plane, suggesting that W3 is likely moving out from the Galactic plane. To confirm this scenario, stellar radial velocity measurements are required to properly evaluate the three velocity components UVW in the Galactic coordinate system.}
    
    The radial velocity of the gas in the direction of W3 is about $-$45\,km\,s$^{-1}$ \citep{Xu06}, which depending on the adopted rotation curve, translates into kinematic distances between 2.9\,kpc \citet{Reid14} and 4.2\,kpc \citet{Russeil03}.     
    \citet{Xu06} reported a difference of $\sim$\,15\,km\,s$^{-1}$ between the rotation velocity of W3(OH) and the velocity expected from Galactic rotation models (e.g. \citealt{Brand93}).
     By combining the velocity components on the other directions, those authors found that the peculiar motion of W3(OH) is about 22\,km\,s$^{-1}$. 
     {The decomposition of the peculiar motion lead to a tangential velocity of about 15\,km\,s$^{-1}$, which is about twice the tangential velocity of $\sim$\,8\,km\,s$^{-1}$, perpendicular to the Galactic plane, that we found for the whole sample of stars from the W3\,Complex (see Sect.\,\ref{sec_3d}).}
     
    W3 is not an exception in terms of peculiar motions or divergence between kinematic and non-kinematic distances. 
    Similar behaviour as that observed in W3(OH) was previously detected in other regions of the outer second quadrant of the Galaxy, where streaming motions with velocities of 15-20\,km\,s$^{-1}$ are well-known \citep{Brand93,Digel96}.
    \citet{Baba09} presents a list of star-forming regions that exhibits peculiar motions with velocities between 20-30\,km\,s$^{-1}$.
    Later, \citet{Moises11} compared the spectrophotometric and kinematic distances of 35 star-forming regions located at different directions in the Galactic plane. Those authors found that about half of their sample is located at a closer distance than their kinematic distances.
     {More recently, \citet{Choi14} presented a large-scale study of the distance and proper-motions of star-forming regions in the Perseus arm leading to the conclusion that, in average, the spiral arm is rotating slower than expected from Galactic rotation models.}
    {In general,} the divergence between the radial velocity predicted by the rotation curves and the observed velocities can be explained by internal processes within the star-forming regions, such as the local stellar winds from high-mass stars \citep{Kudritzki00}, or can be caused by external processes, such as the interaction between supernovae and GMCs (e.g. the HB\,3 and W3, see \citealt{Zhou16}), or fluctuations in the Galactic gravitational potential \citep{Junqueira13}.

\section{Conclusions}
\label{sec_conclusions}

We investigated the {\Gaia} parallaxes of the OB stellar population of the high-density layer of W3, located in the Perseus Arm, to infer the distance {to the complex of {\hii} regions} and its {main} sub-structures. { Based on their astrometric parameters, we selected 37 new objects that are likely associated with W3.}

\begin{enumerate}    
    \item Based on the parallaxes of 68 sources and the Exponentially Decreasing Space Density prior from \citet{Bailer-Jones15}, the distance to W3 was estimated in {2.14$^{+0.08}_{-0.07}$\,kpc}, in agreement with previous distances determined through trigonometric parallaxes of masers or spectrophotometric analysis. The distance to W3 based only on the initial sample of 31 OB stars is {2.23$^{+0.15}_{-0.14}$\,kpc}.
    \item Kinematic distances of the W3 complex are roughly a factor of two larger than the distance obtained from {\Gaia} parallaxes or distances derived from other non-kinematic methods.
    Even when adopting different Galactic rotation curves (e.g. \citealt{Russeil03} and \citealt{Reid14}), the kinematic distances of the W3 complex range between values of 2.9 and 4.2\,kpc, with uncertainties up to 0.7\,kpc.
    \item We further derived the distances to the three main sub-structures of the high-density layer region of W3. {The analysis of their distances based on the parallaxes of the OB stars {tentatively} suggests that they are located at different distances in the line-of-sight. The larger sample of {\Gaia} sources corroborates this hypothesis with relatively greater accuracy:}
    W3\,Main (to the NW) is located at the outer edge of the W3\,Complex ({$d$\,=\,2.30$^{+0.19}_{-0.16}$\,kpc}),
    the W3\,Cluster is at the centre of the W3 complex at a distance of {$d$\,=\,2.17$^{+0.12}_{-0.11}$\,kpc},
    and the W3(OH) is likely located at the closer edge of the complex, at {$d$\,=\,2.00$^{+0.29}_{-0.23}$\,kpc}.
    Combining position of the sub-regions with their distance, the hierarchical scenario of the formation of the high-density layer of W3 is {roughly} consistent with the assumption that the W3\,Cluster triggered the formation of W3\,Main and W3(OH), located in the outer and inner edge of the W3\,Complex, respectively.
    \item {We computed the distance to W3 using the self-gravitating isothermal Galactic disc prior \citep{Apellaniz01} which is specific for the population of OB stars in the Solar neighbourhood, and confirmed our results obtained with the Exponentially Decreasing Space Density prior \citep{Bailer-Jones15}. Both priors return consistent results for the W3\,Complex due to its location within the Galactic disc.} 
\end{enumerate}    

\section*{Acknowledgements}

{The authors thank the anonymous referee who gave very helpful comments and suggestions to improve this work. We thank K.~M.~Menten and M.~Reid for constructive comments and suggestions on this work.
We also thank J.~Ma\'iz-Apellan\'iz for providing his code to compute distances based on the self-gravitating isothermal Galactic disc prior.}
FN thanks the Funda\c{c}\~ao de Amparo \`a Pesquisa do Estado de S\~ao Paulo (FAPESP) for support through process number 2017/18191-8.
AD acknowledges FAPESP (2011/51680-6).

{This work has made use of data from the European Space Agency (ESA) mission {\it Gaia} (\url{https://www.cosmos.esa.int/gaia}), processed by the {\it Gaia} Data Processing and Analysis Consortium (DPAC, \url{https://www.cosmos.esa.int/web/gaia/dpac/consortium}). Funding for the DPAC has been provided by national institutions, in particular the institutions participating in the {\it Gaia} Multilateral Agreement.}


\bibliographystyle{mnras}

\clearpage
\appendix

\section{Comparison between different priors}
\label{appendix_priors}

We compared the distance to the W3\,Complex from Sect.\ref{sec_gaia_w3} assuming the self-gravitating, isothermal Galactic disc prior from \citet{Apellaniz01}, obtained from the study of the distribution of OB stars in the Solar neighbourhood measured by Hipparcos:
\begin{equation}
P(z) = \frac{1-f}{ \cosh^2\left(\frac{z+z_{\odot}}{2 h_{\rm d}} \right) } + f \exp\left( -\frac{1}{2} \left( \frac{z+z_{\odot}}{h_{\rm h}} \right)^2 \right)
    \label{eq:apellaniz}
\end{equation}
\noindent where $z$\,=\,$r\sin(b)$, $b$\,=\,+1.065$^\circ$ is the Galactic latitude of W3, $z_\odot$\,=\,20.0\,pc is the position of the Sun above the Galactic plane, $f$ is the fraction of the halo and the disc stellar populations (we set $f$\,=\,0 since the W3\,Complex is within the Galactic disc), $h_{\rm d}$\,=\,31.8\,pc and $h_{\rm h}$\,=\,490\,pc are the height-scale of the disc and the halo, respectively, taken from \citet{Apellaniz05}. 

\begin{figure}
  \includegraphics[width=\linewidth]{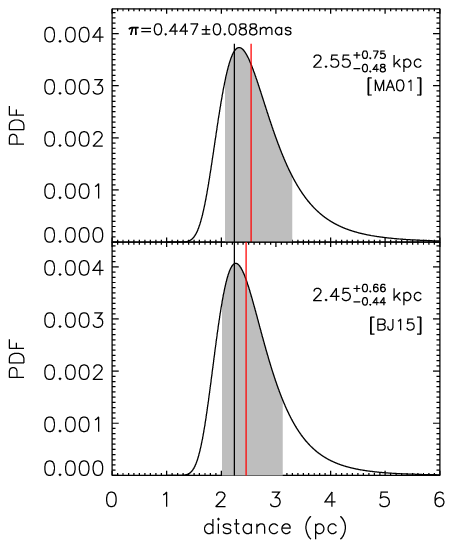}  \\[-6.0ex] 
\caption{{Probability density function (PDF) of the posterior distance estimate based on the mean parallax of the W3\,Complex and using the priors from \citet{Apellaniz01} (top) and \citet{Bailer-Jones15} (bottom). The naive inversion of the mean parallax is indicated by the vertical black line. The median value of the PDF is indicated by the red vertical line. The errors correspond to the 1-$\sigma$ (68\%) confidence interval (shaded grey area)}.}
    \label{fig:pdf_posteriors}
\end{figure}

Figure\,\ref{fig:pdf_posteriors} presents the PDFs of the distance based on both priors and using the mean parallax of the W3\,Complex, $\pi$\,=\,0.437\,$\pm$\,0.088\,mas from Table\,\ref{tab:parallaxes_w3}. The resulting distances and their associated errors are $d$\,=\,2.55$^{+0.75}_{-0.48}$\,kpc and 2.45$^{+0.66}_{-0.43}$\,kpc for the \citet{Apellaniz01} and \citet{Bailer-Jones15} priors, respectively.

We found that both priors led to similar distances, and the usage of one prior or another does not improve the resulting distance to W3.  For this reason, we adopted the EDSD prior from \Bailer in our analysis. 

\bsp	
\label{lastpage}
\end{document}

%% file: table_ob_stars.tex
\setlength{\tabcolsep}{6pt}
\begin{table*}	
  \caption{Known OB stars associated {with the} W3\,Complex.}	
  \label{tab:ob_stars_full}
  {\centering
  \begin{tabular}{r|lccrcc|cc}	
    \hline
\multicolumn{1}{c|}{ID}	&	\multicolumn{1}{c}{Source}	&	\multicolumn{1}{c}{RA}	&	\multicolumn{1}{c}{DEC}	&	\multicolumn{1}{c}{Spectral}	&	\multicolumn{1}{c}{Region}	&	\multicolumn{1}{c}{Ref.}	&	\multicolumn{1}{|c}{2MASS}	&	\multicolumn{1}{c}{Offset}	\\
	&	 	&	\multicolumn{1}{c}{(J2000)}	&	\multicolumn{1}{c}{(J2000)}	&	\multicolumn{1}{c}{Type}	&	&	&	\multicolumn{1}{c}{}	&	\multicolumn{1}{c}{(\arcsec)}	\\
\hline	
1	&	J02251857+6201169	&	02:25:18.6	&	+62:01:17.2	&	O5-O7\,V	&	W3\,Main	&	K15	&	02251857+6201169	&	0.07 	\\
2	&	IRS\,N8	&	02:25:26.3	&	+62:06:07.4	&	YSO	&	Field	&	K15	&	02252633+6206075	&	0.01 	\\
3	&	[NFD2011]\,390	&	02:25:27.4	&	+62:03:43.2	&	B0-2\,V	&	W3\,Main	&	N11	&	02252738+6203432	&	0.11 	\\
4	&	[NFD2011]\,559	&	02:25:28.1	&	+62:05:39.6	&	O7\,V	&	W3\,Main	&	N11	&	02252810+6205395	&	0.22 	\\
5	&	IRS\,4	&	02:25:31.0	&	+62:06:20.6	&	O8-B0.5\,V	&	W3\,Main	&	B12	&	--	&	--	\\
6	&	J02253167+6203249	&	02:25:31.7	&	+62:03:25.2	&	B1\,V	&	W3\,Main	&	K15	&	02253167+6203249	&	0.00 	\\
7	&	IRS\,N2	&	02:25:32.6	&	+62:06:59.8	&	B1-B2\,V	&	W3\,Main	&	B12	&	02253258+6206596	&	0.12 	\\
8	&	IRS\,N5	&	02:25:32.7	&	+62:05:08.1	&	B1-3\,V	&	W3\,Main	&	K15	&	02253274+6205079	&	0.59 	\\
9	&	J02253461+6201401	&	02:25:34.6	&	+62:01:40.3	&	B7\,V	&	W3\,Main	&	K15	&	02253461+6201401	&	0.19 	\\
10	&	IRS\,N1	&	02:25:35.1	&	+62:05:34.5	&	B2-3\,V	&	W3\,Main	&	B12	&	02253517+6205348	&	0.04 	\\
11	&	[NFD2011]\,386	&	02:25:37.5	&	+62:05:24.8	&	B0-B2\,V	&	W3\,Main	&	N11	&	02253750+6205244	&	0.09 	\\
12	&	IRS\,3a	&	02:25:37.8	&	+62:05:51.8	&	O5-7\,V	&	W3\,Main	&	B12	&	02253778+6205522	&	0.04 	\\
13	&	J02253880+6208168	&	02:25:38.8	&	+62:08:17.0	&	B2\,V	&	W3\,Main	&	K15	&	02253880+6208168	&	0.11 	\\
14	&	IRS\,7	&	02:25:40.5	&	+62:05:39.8	&	O9-B2\,V	&	W3\,Main	&	B12	&	--	&	--	\\
15	&	IRS\,N7	&	02:25:40.6	&	+62:05:46.8	&	YSO	&	W3\,Main	&	K15	&	02254062+6205470	&	0.00 	\\
16	&	IRS\,5	&	02:25:40.8	&	+62:05:52.3	&	YSO	&	W3\,Main	&	K15	&	--	&	--	\\
17	&	IRS\,2b	&	02:25:41.7	&	+62:06:24.2	&	B0-1\,V	&	W3\,Main	&	B12	&	--	&	--	\\
18	&	IRS\,2a	&	02:25:43.3	&	+62:06:15.7	&	O8-O9\,V	&	W3\,Main	&	B12	&	02254334+6206154	&	0.08 	\\
19	&	IRS\,2	&	02:25:44.3	&	+62:06:11.4	&	O6.5-7.5\,V	&	W3\,Main	&	B12	&	--	&	--	\\
20	&	[NFD2011]\,347	&	02:25:44.9	&	+62:03:41.5	&	B0-2\,V	&	W3\,Main	&	N11	&	02254485+6203413	&	0.17 	\\
21	&	J02254488+6208155	&	02:25:44.9	&	+62:08:15.8	&	B4-5\,V	&	W3\,Main	&	K15	&	02254488+6208155	&	0.12 	\\
22	&	IRS\,2c	&	02:25:47.1	&	+62:06:13.0	&	B0-1\,V	&	W3\,Main	&	B12	&	02254709+6206131	&	0.06 	\\
23	&	J02254720+6153430	&	02:25:47.2	&	+61:53:43.3	&	B3-5\,V	&	Field	&	K15	&	02254720+6153430	&	0.09 	\\
24	&	IRS\,N6	&	02:25:47.4	&	+62:06:55.3	&	B4\,V	&	W3\,Main	&	K15	&	02254748+6206543	&	0.03 	\\
25	&	J02255220+6156120	&	02:25:52.2	&	+61:56:12.3	&	B:	&	Field	&	B12	&	02255220+6156120	&	0.13 	\\
26	&	J02260587+6158465	&	02:26:05.9	&	+61:58:46.7	&	B1.5\,V	&	W3\,Cluster	&	K15	&	02260587+6158465	&	0.17 	\\
27	&	J02260729+6202550	&	02:26:07.3	&	+62:02:55.1	&	A1\,V	&	Field	&	B12	&	02260729+6202550	&	0.19 	\\
28	&	J02261050+6158018	&	02:26:10.5	&	+61:58:02.0	&	B3\,V	&	W3\,Cluster	&	K15	&	02261050+6158018	&	0.15 	\\
29	&	J02261690+6207350	&	02:26:16.9	&	+62:07:35.2	&	B:	&	Field	&	K15	&	02261690+6207350	&	0.13 	\\
30	&	J02261880+6159529	&	02:26:18.8	&	+61:59:53.1	&	B4\,V	&	W3\,Cluster	&	K15	&	02261880+6159529	&	0.16 	\\
31	&	J02262290+6200370	&	02:26:22.9	&	+62:00:37.3	&	B1-2\,V	&	W3\,Cluster	&	K15	&	02262290+6200370	&	0.09 	\\
32	&	J02262442+6200501	&	02:26:24.4	&	+62:00:50.4	&	B4-6\,V	&	W3\,Cluster	&	K15	&	02262442+6200501	&	0.09 	\\
33	&	J02263433+6201527	&	02:26:34.3	&	+62:01:53.0	&	B4-5\,V	&	W3\,Cluster	&	K15	&	02263433+6201527	&	0.13 	\\
34	&	[NFD2011]\,159	&	02:26:34.4	&	+62:00:42.4	&	O6.5\,V	&	W3\,Cluster	&	N11	&	02263440+6200426	&	0.25 	\\
35	&	[OWK2005]\,1007	&	02:26:41.3	&	+61:59:24.6	&	B2\,V	&	W3\,Cluster	&	K15	&	02264129+6159242	&	0.02 	\\
36	&	J02264173+6158503	&	02:26:41.7	&	+61:58:50.6	&	B4-5\,V	&	W3\,Cluster	&	K15	&	02264173+6158503	&	0.10 	\\
37	&	J02264570+6159504	&	02:26:45.7	&	+61:59:50.7	&	B6-7\,V	&	W3\,Cluster	&	K15	&	02264570+6159504	&	0.07 	\\
38	&	[OWK2005] 4012	&	02:26:46.7	&	+62:00:26.2	&	B2\,V	&	W3\,Cluster	&	K15	&	02264664+6200259	&	0.06 	\\
39	&	J02265627+6206313	&	02:26:56.3	&	+62:06:31.5	&	B0-3\,V	&	Field	&	K15	&	02265627+6206313	&	0.10 	\\
40	&	J02270938+6154437	&	02:27:09.4	&	+61:54:44.0	&	B1-3\,V	&	W3(OH)	&	K15	&	02270938+6154437	&	0.08 	\\
41	&	J02271290+6151390	&	02:27:12.9	&	+61:51:39.2	&	B1-2\,V	&	W3(OH)	&	B12	&	02271290+6151390	&	0.07 	\\
42	&	J02271510+6200151	&	02:27:15.1	&	+62:00:15.3	&	B2\,V	&	Field	&	B12	&	02271510+6200151	&	0.11 	\\
43	&	J02271602+6200506	&	02:27:16.0	&	+62:00:50.8	&	B:	&	Field	&	K15	&	02271602+6200506	&	0.15 	\\
44	&	J02272023+6202023	&	02:27:20.3	&	+62:02:02.6	&	B0-1.5\,V	&	Field	&	K15	&	02272023+6202023	&	0.05 	\\
45	&	[NFD2011]\,252	&	02:27:21.4	&	+61:54:57.3	&	B1-B4 V	&	W3(OH)	&	N11	&	02272133+6154570	&	0.04 	\\
46	&	[NFD2011]\,3	&	02:27:34.6	&	+61:55:57.2	&	B0-1IIIe	&	W3(OH)	&	N11	&	02273459+6155571	&	0.06 	\\
47	&	J02273953+6156330	&	02:27:39.6	&	+61:56:33.3	&	B0-B2 V	&	W3(OH)	&	K15	&	02273953+6156330	&	0.06 	\\
48	&	J02274017+6204197	&	02:27:40.2	&	+62:04:20.0	&	B1-4\,V	&	Field	&	K15	&	02274017+6204197	&	0.07 \\
    \hline	
  \end{tabular}
  }\\
{\textbf{Notes:} {the columns are as follows:
(1) ID of the source;
(2) identification of the OB star based on the name given in the referenced catalogues;
(3) right ascension;
(4) declination;
(5) spectral type; 
(6) association with a sub-region of the W3\,Complex;
(7) references for the properties of the OB stars: N11 -- \citealt{Navarete11}; B12 -- \citealt{Bik12}; K15 -- \citealt{Kiminki15};
(8) 2MASS counterpart;
(9) offset between the {\Gaia} and the 2MASS counterpart.}}
\end{table*}	
\setlength{\tabcolsep}{6pt}

%% file: table_gaia.tex
\setlength{\tabcolsep}{3pt}	
 \begin{table*}	
	\centering	
	\caption{{{\Gaia} information of the OB-type stellar population of the W3\,Complex.}}
	\label{tab:gaia_parallaxes}	
	\begin{tabular}{rr|cccrrcr|cccc} 	
\hline
\multicolumn{1}{c}{ID}	&	\multicolumn{1}{c|}{Designation}	&	\multicolumn{1}{c}{sep}	&	\multicolumn{1}{c}{$\alpha$}	&	\multicolumn{1}{c}{$\delta$}	&	\multicolumn{1}{c}{$\mu_\alpha \cos(\delta)$}	&	\multicolumn{1}{c}{$\mu_\delta$}	&	\multicolumn{1}{c}{$\mu$}	&	\multicolumn{1}{c|}{$\pi$}	&	\multicolumn{1}{c}{G}	&	\multicolumn{1}{c}{[BP$-$RP]}	&	\multicolumn{1}{c}{RUWE}	&	\multicolumn{1}{c}{ }	\\
	&	\multicolumn{1}{c|}{({\it Gaia} DR2 \#)}	&	\multicolumn{1}{c}{(\arcsec)}	&	\multicolumn{1}{c}{(deg)}	&	\multicolumn{1}{c}{(deg)}	&	\multicolumn{1}{c}{(mas\,yr$^{-1}$)}	&	\multicolumn{1}{c}{(mas\,yr$^{-1}$)}	&	\multicolumn{1}{c}{(mas\,yr$^{-1}$)}	&	\multicolumn{1}{c|}{(mas)}	&	\multicolumn{1}{c}{(mag)}	&	\multicolumn{1}{c}{(mag)}	&	\multicolumn{1}{c}{ }	&	\multicolumn{1}{c}{ }	\\
\hline	
1	&	513638063807365248	&	0.07 	&	36.327350 	&	62.021349 	&	$-$1.388\,$\pm$\,0.071 	&	$-$1.817\,$\pm$\,0.091 	&	2.29\,$\pm$\,0.12 	&	0.423\,$\pm$\,0.062 	&	14.21 	&	3.60 	&	1.11 	&	\\
3	&	513638579203426816	&	0.11 	&	36.364035 	&	62.061969 	&	$-$2.380\,$\pm$\,0.056 	&	$-$2.611\,$\pm$\,0.071 	&	3.53\,$\pm$\,0.09 	&	0.379\,$\pm$\,0.048 	&	15.02 	&	2.96 	&	1.08 	&	\\
4	&	513638922800802048	&	0.22 	&	36.367054 	&	62.094276 	&	1.294\,$\pm$\,0.453 	&	$-$0.520\,$\pm$\,0.546 	&	1.39\,$\pm$\,0.71 	&	$-$1.890\,$\pm$\,0.384 	&	13.19 	&	2.91 	&	8.56 	&	$\ast$	\\
6	&	513638510483951232	&	0.07 	&	36.381921 	&	62.056924 	&	$-$0.983\,$\pm$\,0.029 	&	$-$1.035\,$\pm$\,0.037 	&	1.43\,$\pm$\,0.05 	&	0.452\,$\pm$\,0.024 	&	13.82 	&	2.23 	&	1.03 	&	\\
7	&	513638957160533888	&	0.12 	&	36.385753 	&	62.116530 	&	0.728\,$\pm$\,0.389 	&	$-$2.747\,$\pm$\,0.480 	&	2.84\,$\pm$\,0.62 	&	$-$0.704\,$\pm$\,0.343 	&	16.29 	&	3.83 	&	3.90 	&	$\ast$	\\
9	&	513637376612594688	&	0.19 	&	36.394190 	&	62.027765 	&	$-$0.797\,$\pm$\,0.044 	&	$-$0.916\,$\pm$\,0.055 	&	1.21\,$\pm$\,0.07 	&	0.366\,$\pm$\,0.038 	&	15.78 	&	1.92 	&	1.03 	&	\\
11	&	513638751002110336	&	0.09 	&	36.406259 	&	62.090096 	&	$-$0.508\,$\pm$\,0.171 	&	$-$1.647\,$\pm$\,0.197 	&	1.72\,$\pm$\,0.26 	&	0.420\,$\pm$\,0.143 	&	17.40 	&	4.15 	&	1.15 	&	\\
13	&	513662459222193536	&	0.11 	&	36.411647 	&	62.137995 	&	$-$1.168\,$\pm$\,0.052 	&	$-$0.961\,$\pm$\,0.067 	&	1.51\,$\pm$\,0.08 	&	0.374\,$\pm$\,0.046 	&	15.34 	&	2.37 	&	1.21 	&	\\
18	&	513662218697964928	&	0.08 	&	36.430548 	&	62.104269 	&	0.250\,$\pm$\,0.172 	&	$-$3.855\,$\pm$\,0.200 	&	3.86\,$\pm$\,0.26 	&	0.687\,$\pm$\,0.156 	&	17.49 	&	3.39 	&	1.34 	&	\\
20	&	513637892008658048	&	0.17 	&	36.436857 	&	62.061437 	&	$-$0.196\,$\pm$\,0.061 	&	$-$2.212\,$\pm$\,0.077 	&	2.22\,$\pm$\,0.10 	&	0.413\,$\pm$\,0.054 	&	14.17 	&	3.23 	&	1.08 	&	\\
21	&	513662562301408384	&	0.12 	&	36.436991 	&	62.137638 	&	$-$0.639\,$\pm$\,0.043 	&	$-$0.061\,$\pm$\,0.056 	&	0.64\,$\pm$\,0.07 	&	0.413\,$\pm$\,0.037 	&	15.74 	&	2.00 	&	1.01 	&	\\
22	&	513662149984550912	&	0.06 	&	36.446185 	&	62.103628 	&	$-$0.331\,$\pm$\,0.794 	&	$-$2.509\,$\pm$\,0.977 	&	2.53\,$\pm$\,1.26 	&	0.973\,$\pm$\,0.662 	&	20.19 	&	3.41 	&	1.08 	&	$\ast$	\\
23	&	513632119572654848	&	0.09 	&	36.446698 	&	61.895264 	&	$-$0.575\,$\pm$\,0.055 	&	$-$0.454\,$\pm$\,0.073 	&	0.73\,$\pm$\,0.09 	&	0.346\,$\pm$\,0.052 	&	15.98 	&	2.31 	&	1.05 	&	\\
24	&	513662180045929728	&	0.25 	&	36.447991 	&	62.115128 	&	$-$0.773\,$\pm$\,0.185 	&	$-$0.648\,$\pm$\,0.220 	&	1.01\,$\pm$\,0.29 	&	0.472\,$\pm$\,0.161 	&	17.88 	&	3.63 	&	1.16 	&	\\
25	&	513633871919302016	&	0.13 	&	36.467507 	&	61.936648 	&	$-$1.369\,$\pm$\,0.039 	&	$-$0.323\,$\pm$\,0.052 	&	1.41\,$\pm$\,0.06 	&	0.399\,$\pm$\,0.034 	&	15.60 	&	1.93 	&	0.96 	&	\\
26	&	513634387315470848	&	0.17 	&	36.524411 	&	61.979552 	&	$-$1.030\,$\pm$\,0.034 	&	$-$0.010\,$\pm$\,0.048 	&	1.03\,$\pm$\,0.06 	&	0.439\,$\pm$\,0.029 	&	14.14 	&	2.10 	&	1.21 	&	\\
27	&	513661050473012992	&	0.19 	&	36.530359 	&	62.048567 	&	$-$0.557\,$\pm$\,0.033 	&	0.126\,$\pm$\,0.046 	&	0.57\,$\pm$\,0.06 	&	0.432\,$\pm$\,0.030 	&	15.17 	&	1.56 	&	1.02 	&	\\
28	&	513634181157041792	&	0.15 	&	36.543749 	&	61.967146 	&	$-$0.312\,$\pm$\,0.044 	&	$-$0.626\,$\pm$\,0.063 	&	0.70\,$\pm$\,0.08 	&	0.511\,$\pm$\,0.039 	&	15.82 	&	2.18 	&	0.96 	&	\\
29	&	513661905166430080	&	0.13 	&	36.570399 	&	62.126359 	&	$-$0.648\,$\pm$\,0.085 	&	$-$0.613\,$\pm$\,0.114 	&	0.89\,$\pm$\,0.14 	&	0.491\,$\pm$\,0.079 	&	15.76 	&	3.37 	&	1.12 	&	\\
30	&	513634352955730816	&	0.16 	&	36.578321 	&	61.998003 	&	$-$0.652\,$\pm$\,0.027 	&	0.101\,$\pm$\,0.040 	&	0.66\,$\pm$\,0.05 	&	0.468\,$\pm$\,0.025 	&	14.67 	&	1.61 	&	1.01 	&	\\
31	&	513657820657620864	&	0.08 	&	36.595387 	&	62.010258 	&	$-$0.694\,$\pm$\,0.035 	&	$-$0.487\,$\pm$\,0.054 	&	0.85\,$\pm$\,0.06 	&	0.372\,$\pm$\,0.033 	&	14.38 	&	2.37 	&	1.03 	&	\\
32	&	513657820657620736	&	0.09 	&	36.601735 	&	62.013918 	&	$-$0.807\,$\pm$\,0.073 	&	$-$0.357\,$\pm$\,0.112 	&	0.88\,$\pm$\,0.13 	&	0.381\,$\pm$\,0.067 	&	15.99 	&	2.88 	&	1.12 	&	\\
33	&	513657889377083520	&	0.13 	&	36.642966 	&	62.031299 	&	$-$2.307\,$\pm$\,0.098 	&	1.795\,$\pm$\,0.127 	&	2.92\,$\pm$\,0.16 	&	0.738\,$\pm$\,0.090 	&	14.03 	&	1.35 	&	4.39 	&	$\ast$	\\
34	&	513657786292622848	&	0.28 	&	36.643312 	&	62.011764 	&	$-$0.455\,$\pm$\,0.071 	&	0.575\,$\pm$\,0.087 	&	0.73\,$\pm$\,0.11 	&	0.393\,$\pm$\,0.057 	&	9.89 	&	1.43 	&	1.87 	&	$\ast$	\\
35	&	513656961664153728	&	0.02 	&	36.672069 	&	61.990061 	&	$-$0.645\,$\pm$\,0.022 	&	$-$0.090\,$\pm$\,0.034 	&	0.65\,$\pm$\,0.04 	&	0.452\,$\pm$\,0.021 	&	13.88 	&	1.40 	&	1.01 	&	\\
37	&	513657064743367552	&	0.07 	&	36.690397 	&	61.997331 	&	$-$0.669\,$\pm$\,0.059 	&	$-$0.636\,$\pm$\,0.084 	&	0.92\,$\pm$\,0.10 	&	0.304\,$\pm$\,0.053 	&	15.89 	&	2.21 	&	1.02 	&	\\
38	&	513657064743365760	&	0.06 	&	36.694330 	&	62.007184 	&	$-$0.522\,$\pm$\,0.021 	&	0.094\,$\pm$\,0.031 	&	0.53\,$\pm$\,0.04 	&	0.441\,$\pm$\,0.019 	&	13.51 	&	1.44 	&	0.93 	&	\\
39	&	513660019680852992	&	0.10 	&	36.734476 	&	62.108671 	&	$-$0.655\,$\pm$\,0.048 	&	$-$0.692\,$\pm$\,0.067 	&	0.95\,$\pm$\,0.08 	&	0.479\,$\pm$\,0.043 	&	14.90 	&	2.68 	&	1.07 	&	\\
40	&	513609373425115136	&	0.07 	&	36.789058 	&	61.912138 	&	$-$0.979\,$\pm$\,0.061 	&	$-$0.070\,$\pm$\,0.089 	&	0.98\,$\pm$\,0.11 	&	0.456\,$\pm$\,0.060 	&	16.22 	&	2.67 	&	0.96 	&	\\
41	&	513585046729657216	&	0.07 	&	36.803742 	&	61.860829 	&	$-$1.339\,$\pm$\,0.030 	&	$-$0.031\,$\pm$\,0.044 	&	1.34\,$\pm$\,0.05 	&	0.485\,$\pm$\,0.028 	&	13.66 	&	2.19 	&	1.11 	&	\\
43	&	513610404217453312	&	0.15 	&	36.816736 	&	62.014023 	&	$-$0.317\,$\pm$\,0.070 	&	$-$0.571\,$\pm$\,0.098 	&	0.65\,$\pm$\,0.12 	&	0.615\,$\pm$\,0.061 	&	14.40 	&	3.29 	&	1.13 	&	\\
44	&	513610438576963712	&	0.05 	&	36.834264 	&	62.033991 	&	$-$0.113\,$\pm$\,0.051 	&	0.245\,$\pm$\,0.070 	&	0.27\,$\pm$\,0.09 	&	0.510\,$\pm$\,0.047 	&	14.81 	&	2.67 	&	1.13 	&	\\
45	&	513608754949821952	&	0.04 	&	36.838890 	&	61.915831 	&	$-$0.867\,$\pm$\,0.069 	&	$-$0.506\,$\pm$\,0.099 	&	1.00\,$\pm$\,0.12 	&	0.465\,$\pm$\,0.062 	&	15.66 	&	3.39 	&	0.99 	&	\\
46	&	513609098547198720	&	0.06 	&	36.894142 	&	61.932525 	&	$-$0.986\,$\pm$\,0.091 	&	0.165\,$\pm$\,0.131 	&	1.00\,$\pm$\,0.16 	&	0.569\,$\pm$\,0.083 	&	14.27 	&	3.54 	&	1.37 	&	\\
47	&	513609132906935168	&	0.06 	&	36.914674 	&	61.942501 	&	$-$0.726\,$\pm$\,0.074 	&	$-$0.589\,$\pm$\,0.103 	&	0.93\,$\pm$\,0.13 	&	0.546\,$\pm$\,0.065 	&	15.78 	&	3.15 	&	1.06 	&	\\
48	&	513612397082030720	&	0.07 	&	36.917376 	&	62.072145 	&	$-$0.273\,$\pm$\,0.058 	&	$-$0.501\,$\pm$\,0.080 	&	0.57\,$\pm$\,0.10 	&	0.500\,$\pm$\,0.050 	&	16.10 	&	2.33 	&	1.03 	&	\\
    \hline
\end{tabular} \\	
    {{\bf Notes:} The columns are as follows:	
    (1) ID from Table\,\ref{tab:ob_stars_full}; 	
    (2) Designation from the {\Gaia} DR2 catalogue;	
    (3) angular separation between the 2MASS and the {\Gaia} coordinates;	
    (4) right ascension of the {\Gaia} source (in decimal degrees);	
    (5) declination of the {\Gaia} Source (in decimal degrees);	
    (6) proper motion on the $\alpha$-axis and its error;	
    (7) proper motion on the $\delta$-axis and its error; 	
    (8) total proper motion and its error;	
    (9) parallax and its error {from the {\Gaia}-DR2 archive};	
    (10) $G$-band magnitude;	
    (11) $BP$-$RP$ colour index;	
    (12) Re-normalised Unit Weight Error (RUWE);	
    (13) sources flagged with an $\ast$ were removed from the analysis {based on the RUWE selection criterion (RUWE\,$\leq$\,1.4) to remove sources with poor astrometry}.	
}	
\end{table*}	
\setlength{\tabcolsep}{6pt}

%% file: table_gaia_newsample.tex
\setlength{\tabcolsep}{3pt}	
 \begin{table*}	
	\caption{{{\Gaia} data of the new W3 members selected in this study.}}
	\label{tab:gaia_parallaxes_largersample}	
	\begin{tabular}{rrc|ccrrcr|ccc} 	
\hline
\multicolumn{1}{c}{ID}	&	\multicolumn{1}{c}{Designation}	&	\multicolumn{1}{c|}{W3}	&	\multicolumn{1}{c}{$\alpha$}	&	\multicolumn{1}{c}{$\delta$}	&	\multicolumn{1}{c}{$\mu_\alpha \cos(\delta)$}	&	\multicolumn{1}{c}{$\mu_\delta$}	&	\multicolumn{1}{c}{$\mu$}	&	\multicolumn{1}{c|}{$\pi$}	&	\multicolumn{1}{c}{G}	&	\multicolumn{1}{c}{[BP$-$RP]}	&	\multicolumn{1}{c}{RUWE}	\\
	&	\multicolumn{1}{c}{({\it Gaia} DR2 \#)}	&	\multicolumn{1}{c|}{region} &	\multicolumn{1}{c}{(deg)}	&	\multicolumn{1}{c}{(deg)}	&	\multicolumn{1}{c}{(mas\,yr$^{-1}$)}	&	\multicolumn{1}{c}{(mas\,yr$^{-1}$)}	&	\multicolumn{1}{c}{(mas\,yr$^{-1}$)}	&	\multicolumn{1}{c|}{(mas)}	&	\multicolumn{1}{c}{(mag)}	&	\multicolumn{1}{c}{(mag)}	&	\multicolumn{1}{c}{ }		\\
\hline	
G01	&	513586146241283072	&	d	&	36.715834 	&	61.905581 	&	$-$0.637\,$\pm$\,0.034 	&	$-$0.644\,$\pm$\,0.048 	&	0.91\,$\pm$\,0.06 	&	0.406\,$\pm$\,0.031 	&	12.81 	&	1.46 	&	0.95 	\\
G02	&	513608273913499008	&	d	&	36.881812 	&	61.867867 	&	$-$0.950\,$\pm$\,0.123 	&	$-$0.747\,$\pm$\,0.158 	&	1.21\,$\pm$\,0.20 	&	0.458\,$\pm$\,0.109 	&	17.40 	&	2.63 	&	1.05 	\\
G03	&	513609266046952192	&	b	&	36.936479 	&	61.969724 	&	$-$1.029\,$\pm$\,0.133 	&	$-$0.572\,$\pm$\,0.158 	&	1.18\,$\pm$\,0.21 	&	0.537\,$\pm$\,0.110 	&	17.59 	&	2.57 	&	1.06 	\\
G04	&	513610301138022528	&	d	&	36.821306 	&	62.002882 	&	$-$0.683\,$\pm$\,0.035 	&	$-$0.469\,$\pm$\,0.052 	&	0.83\,$\pm$\,0.06 	&	0.505\,$\pm$\,0.032 	&	15.24 	&	1.65 	&	1.17 	\\
G05	&	513630775244268288	&	d	&	36.440729 	&	61.841246 	&	$-$0.722\,$\pm$\,0.206 	&	$-$0.352\,$\pm$\,0.293 	&	0.80\,$\pm$\,0.36 	&	0.362\,$\pm$\,0.191 	&	18.41 	&	3.67 	&	1.04 	\\
G06	&	513631157500074752	&	d	&	36.452212 	&	61.850967 	&	$-$0.737\,$\pm$\,0.083 	&	$-$0.542\,$\pm$\,0.120 	&	0.92\,$\pm$\,0.15 	&	0.481\,$\pm$\,0.078 	&	16.95 	&	2.48 	&	0.99 	\\
G07	&	513632978566202624	&	d	&	36.571849 	&	61.935235 	&	$-$0.647\,$\pm$\,0.147 	&	$-$0.711\,$\pm$\,0.181 	&	0.96\,$\pm$\,0.23 	&	0.514\,$\pm$\,0.125 	&	17.69 	&	2.81 	&	1.05 	\\
G08	&	513633047285680000	&	d	&	36.668163 	&	61.918736 	&	$-$0.885\,$\pm$\,0.327 	&	$-$0.803\,$\pm$\,0.409 	&	1.20\,$\pm$\,0.52 	&	0.443\,$\pm$\,0.261 	&	19.04 	&	2.96 	&	0.99 	\\
G09	&	513633111711389824	&	d	&	36.640956 	&	61.932075 	&	$-$0.829\,$\pm$\,0.192 	&	$-$0.598\,$\pm$\,0.247 	&	1.02\,$\pm$\,0.31 	&	0.369\,$\pm$\,0.176 	&	18.45 	&	2.82 	&	0.99 	\\
G10	&	513633146067364736	&	d	&	36.668250 	&	61.931271 	&	$-$1.041\,$\pm$\,0.642 	&	$-$0.600\,$\pm$\,0.763 	&	1.20\,$\pm$\,1.00 	&	0.551\,$\pm$\,0.579 	&	20.01 	&	2.36 	&	1.07 	\\
G11	&	513633356523322624	&	c	&	36.585116 	&	61.950644 	&	$-$0.767\,$\pm$\,0.187 	&	$-$0.255\,$\pm$\,0.255 	&	0.81\,$\pm$\,0.32 	&	0.496\,$\pm$\,0.168 	&	18.27 	&	2.74 	&	1.08 	\\
G12	&	513633425242797568	&	b	&	36.625383 	&	61.960912 	&	$-$0.917\,$\pm$\,0.206 	&	$-$0.354\,$\pm$\,0.281 	&	0.98\,$\pm$\,0.35 	&	0.415\,$\pm$\,0.179 	&	18.08 	&	4.00 	&	1.07 	\\
G13	&	513633455302772864	&	b	&	36.653876 	&	61.973686 	&	$-$0.735\,$\pm$\,0.383 	&	$-$0.540\,$\pm$\,0.441 	&	0.91\,$\pm$\,0.58 	&	0.408\,$\pm$\,0.311 	&	19.27 	&	2.50 	&	1.02 	\\
G14	&	513633459597914112	&	b	&	36.649997 	&	61.967404 	&	$-$0.715\,$\pm$\,0.313 	&	$-$0.748\,$\pm$\,0.400 	&	1.03\,$\pm$\,0.51 	&	0.508\,$\pm$\,0.274 	&	19.00 	&	2.89 	&	1.02 	\\
G15	&	513633459602534400	&	b	&	36.643851 	&	61.971339 	&	$-$0.774\,$\pm$\,0.078 	&	$-$0.350\,$\pm$\,0.108 	&	0.85\,$\pm$\,0.13 	&	0.407\,$\pm$\,0.070 	&	16.90 	&	2.58 	&	0.93 	\\
G16	&	513633489662505472	&	b	&	36.619071 	&	61.968168 	&	$-$0.696\,$\pm$\,0.227 	&	$-$0.603\,$\pm$\,0.303 	&	0.92\,$\pm$\,0.38 	&	0.474\,$\pm$\,0.201 	&	18.55 	&	2.61 	&	1.10 	\\
G17	&	513633493962273152	&	b	&	36.613435 	&	61.970601 	&	$-$0.840\,$\pm$\,0.058 	&	$-$0.402\,$\pm$\,0.083 	&	0.93\,$\pm$\,0.10 	&	0.357\,$\pm$\,0.052 	&	16.05 	&	2.60 	&	1.02 	\\
G18	&	513633597036933504	&	b	&	36.507163 	&	61.917826 	&	$-$0.841\,$\pm$\,0.441 	&	$-$0.501\,$\pm$\,0.567 	&	0.98\,$\pm$\,0.72 	&	0.491\,$\pm$\,0.352 	&	19.51 	&	2.92 	&	0.99 	\\
G19	&	513633665760970752	&	b	&	36.493734 	&	61.932811 	&	$-$0.889\,$\pm$\,0.126 	&	$-$0.473\,$\pm$\,0.154 	&	1.01\,$\pm$\,0.20 	&	0.354\,$\pm$\,0.106 	&	17.56 	&	2.50 	&	1.00 	\\
G20	&	513634387315471232	&	b	&	36.518742 	&	61.974081 	&	$-$0.982\,$\pm$\,0.056 	&	$-$0.392\,$\pm$\,0.078 	&	1.06\,$\pm$\,0.10 	&	0.468\,$\pm$\,0.049 	&	15.63 	&	2.79 	&	0.99 	\\
G21	&	513634520456485632	&	b	&	36.549796 	&	62.003578 	&	$-$0.917\,$\pm$\,0.210 	&	$-$0.226\,$\pm$\,0.285 	&	0.94\,$\pm$\,0.35 	&	0.366\,$\pm$\,0.183 	&	18.47 	&	2.54 	&	1.03 	\\
G22	&	513637307893119232	&	a	&	36.402893 	&	62.021857 	&	$-$0.591\,$\pm$\,0.039 	&	$-$0.462\,$\pm$\,0.049 	&	0.75\,$\pm$\,0.06 	&	0.464\,$\pm$\,0.034 	&	15.61 	&	1.81 	&	0.97 	\\
G23	&	513637410972338432	&	d	&	36.462507 	&	61.998311 	&	$-$0.807\,$\pm$\,0.056 	&	$-$0.767\,$\pm$\,0.070 	&	1.11\,$\pm$\,0.09 	&	0.453\,$\pm$\,0.049 	&	15.71 	&	2.59 	&	1.07 	\\
G24	&	513637960728155264	&	d	&	36.312037 	&	62.000456 	&	$-$0.692\,$\pm$\,0.109 	&	$-$0.664\,$\pm$\,0.129 	&	0.96\,$\pm$\,0.17 	&	0.356\,$\pm$\,0.096 	&	17.40 	&	2.57 	&	1.00 	\\
G25	&	513638510483950976	&	a	&	36.385138 	&	62.058260 	&	$-$0.700\,$\pm$\,0.054 	&	$-$0.722\,$\pm$\,0.066 	&	1.01\,$\pm$\,0.09 	&	0.479\,$\pm$\,0.046 	&	16.22 	&	2.07 	&	0.99 	\\
G26	&	513638510483952256	&	a	&	36.375288 	&	62.056131 	&	$-$0.677\,$\pm$\,0.112 	&	$-$0.722\,$\pm$\,0.131 	&	0.99\,$\pm$\,0.17 	&	0.458\,$\pm$\,0.093 	&	17.42 	&	2.68 	&	1.00 	\\
G27	&	513650639471583616	&	a	&	36.311688 	&	62.105059 	&	$-$0.723\,$\pm$\,0.173 	&	$-$0.753\,$\pm$\,0.213 	&	1.04\,$\pm$\,0.27 	&	0.343\,$\pm$\,0.150 	&	18.20 	&	3.02 	&	1.02 	\\
G28	&	513656996023891584	&	b	&	36.647729 	&	61.995069 	&	$-$0.689\,$\pm$\,0.472 	&	$-$0.326\,$\pm$\,0.595 	&	0.76\,$\pm$\,0.76 	&	0.551\,$\pm$\,0.377 	&	19.41 	&	2.63 	&	1.07 	\\
G29	&	513656996023892864	&	b	&	36.642495 	&	61.988897 	&	$-$0.638\,$\pm$\,0.132 	&	$-$0.582\,$\pm$\,0.169 	&	0.86\,$\pm$\,0.21 	&	0.342\,$\pm$\,0.116 	&	17.49 	&	2.63 	&	1.01 	\\
G30	&	513657958096560512	&	b	&	36.630819 	&	62.031292 	&	$-$0.530\,$\pm$\,0.600 	&	$-$0.223\,$\pm$\,0.540 	&	0.57\,$\pm$\,0.81 	&	0.429\,$\pm$\,0.349 	&	19.11 	&	2.03 	&	1.07 	\\
G31	&	513658473492629248	&	d	&	36.728785 	&	62.092657 	&	$-$0.812\,$\pm$\,0.090 	&	$-$0.656\,$\pm$\,0.121 	&	1.04\,$\pm$\,0.15 	&	0.541\,$\pm$\,0.081 	&	16.92 	&	2.22 	&	1.03 	\\
G32	&	513659435565297920	&	d	&	36.915298 	&	62.107850 	&	$-$0.586\,$\pm$\,0.043 	&	$-$0.381\,$\pm$\,0.057 	&	0.70\,$\pm$\,0.07 	&	0.511\,$\pm$\,0.038 	&	15.68 	&	1.52 	&	1.05 	\\
G33	&	513659572998812672	&	d	&	36.948379 	&	62.132059 	&	$-$0.594\,$\pm$\,0.624 	&	$-$0.561\,$\pm$\,0.664 	&	0.82\,$\pm$\,0.91 	&	0.409\,$\pm$\,0.473 	&	19.86 	&	2.31 	&	0.98 	\\
G34	&	513660122754670976	&	d	&	36.788270 	&	62.123131 	&	$-$0.800\,$\pm$\,0.160 	&	$-$0.824\,$\pm$\,0.234 	&	1.15\,$\pm$\,0.28 	&	0.516\,$\pm$\,0.149 	&	18.04 	&	2.45 	&	1.03 	\\
G35	&	513660225833910144	&	d	&	36.693976 	&	62.116079 	&	$-$0.869\,$\pm$\,0.123 	&	$-$0.299\,$\pm$\,0.148 	&	0.92\,$\pm$\,0.19 	&	0.531\,$\pm$\,0.104 	&	17.56 	&	2.31 	&	0.98 	\\
G36	&	513660363278230272	&	d	&	36.715726 	&	62.147274 	&	$-$0.729\,$\pm$\,0.110 	&	$-$0.318\,$\pm$\,0.132 	&	0.80\,$\pm$\,0.17 	&	0.540\,$\pm$\,0.092 	&	17.42 	&	1.59 	&	1.03 	\\
G37	&	513662631020884608	&	a	&	36.406787 	&	62.149860 	&	$-$0.799\,$\pm$\,0.129 	&	$-$0.473\,$\pm$\,0.150 	&	0.93\,$\pm$\,0.20 	&	0.397\,$\pm$\,0.113 	&	17.40 	&	3.16 	&	1.00 	\\

 \hline
\end{tabular} \\	
    {{\bf Notes:} {The columns are as follows:	
    (1) ID of the Source; 	
    (2) Designation from the {\Gaia}-DR2 catalogue;	
    (3) Association with a sub-region of the W3\,Complex: (a) W3\,Main, (b) W3\,Cluster, (c) W3(OH), or (d) field star.
    (3) right ascension of the {\Gaia} source (in decimal degrees);		
    (4) declination of the {\Gaia} Source (in decimal degrees);	
    (5) proper motion on the $\alpha$-axis and its error;	
    (6) proper motion on the $\delta$-axis and its error; 
    (7)	total proper motion and its error;	
    (8) parallax and its error from the {\Gaia}-DR2 archive;
    (9) $G$-band magnitude;	
    (10) $BP$-$RP$ colour index;	
    (11) Re-normalised Unit Weight Error (RUWE).}	
}	
\end{table*}	
\setlength{\tabcolsep}{6pt}

%% file: MN-19-0699-MJ_final.bbl
\begin{thebibliography}{}
\makeatletter
\relax
\def\mn@urlcharsother{\let\do\@makeother \do\$\do\&\do\#\do\^\do\_\do\%\do\~}
\def\mn@doi{\begingroup\mn@urlcharsother \@ifnextchar [ {\mn@doi@}
  {\mn@doi@[]}}
\def\mn@doi@[#1]#2{\def\@tempa{#1}\ifx\@tempa\@empty \href
  {http://dx.doi.org/#2} {doi:#2}\else \href {http://dx.doi.org/#2} {#1}\fi
  \endgroup}
\def\mn@eprint#1#2{\mn@eprint@#1:#2::\@nil}
\def\mn@eprint@arXiv#1{\href {http://arxiv.org/abs/#1} {{\tt arXiv:#1}}}
\def\mn@eprint@dblp#1{\href {http://dblp.uni-trier.de/rec/bibtex/#1.xml}
  {dblp:#1}}
\def\mn@eprint@#1:#2:#3:#4\@nil{\def\@tempa {#1}\def\@tempb {#2}\def\@tempc
  {#3}\ifx \@tempc \@empty \let \@tempc \@tempb \let \@tempb \@tempa \fi \ifx
  \@tempb \@empty \def\@tempb {arXiv}\fi \@ifundefined
  {mn@eprint@\@tempb}{\@tempb:\@tempc}{\expandafter \expandafter \csname
  mn@eprint@\@tempb\endcsname \expandafter{\@tempc}}}

\bibitem[\protect\citeauthoryear{{Astraatmadja} \&
  {Bailer-Jones}}{{Astraatmadja} \& {Bailer-Jones}}{2016}]{Astraatmadja16}
{Astraatmadja} T.~L.,  {Bailer-Jones} C.~A.~L.,  2016, \mn@doi [\apj]
  {10.3847/1538-4357/833/1/119}, \href
  {http://adsabs.harvard.edu/abs/2016ApJ...833..119A} {833, 119}

\bibitem[\protect\citeauthoryear{{Baba}, {Asaki}, {Makino}, {Miyoshi}, {Saitoh}
   \& {Wada}}{{Baba} et~al.}{2009}]{Baba09}
{Baba} J.,  {Asaki} Y.,  {Makino} J.,  {Miyoshi} M.,  {Saitoh} T.~R.,   {Wada}
  K.,  2009, \mn@doi [\apj] {10.1088/0004-637X/706/1/471}, \href
  {http://adsabs.harvard.edu/abs/2009ApJ...706..471B} {706, 471}

\bibitem[\protect\citeauthoryear{{Bailer-Jones}}{{Bailer-Jones}}{2015}]{Bailer-Jones15}
{Bailer-Jones} C.~A.~L.,  2015, \mn@doi [\pasp] {10.1086/683116}, \href
  {http://adsabs.harvard.edu/abs/2015PASP..127..994B} {127, 994}

\bibitem[\protect\citeauthoryear{{Becker} \& {Fenkart}}{{Becker} \&
  {Fenkart}}{1971}]{Becker71}
{Becker} W.,  {Fenkart} R.,  1971, \aaps, \href
  {http://adsabs.harvard.edu/abs/1971A%26AS....4..241B} {4, 241}

\bibitem[\protect\citeauthoryear{{Bik} et~al.,}{{Bik} et~al.}{2012}]{Bik12}
{Bik} A.,  et~al., 2012, \mn@doi [\apj] {10.1088/0004-637X/744/2/87}, \href
  {http://adsabs.harvard.edu/abs/2012ApJ...744...87B} {744, 87}

\bibitem[\protect\citeauthoryear{{Blum}, {Conti}  \& {Damineli}}{{Blum}
  et~al.}{2000}]{Blum00}
{Blum} R.~D.,  {Conti} P.~S.,   {Damineli} A.,  2000, \mn@doi [\aj]
  {10.1086/301317}, \href {http://adsabs.harvard.edu/abs/2000AJ....119.1860B}
  {119, 1860}

\bibitem[\protect\citeauthoryear{{Brand} \& {Blitz}}{{Brand} \&
  {Blitz}}{1993}]{Brand93}
{Brand} J.,  {Blitz} L.,  1993, \aap, \href
  {http://adsabs.harvard.edu/abs/1993A%26A...275...67B} {275, 67}

\bibitem[\protect\citeauthoryear{{Cardelli}, {Clayton}  \& {Mathis}}{{Cardelli}
  et~al.}{1989}]{Cardelli89}
{Cardelli} J.~A.,  {Clayton} G.~C.,   {Mathis} J.~S.,  1989, \mn@doi [\apj]
  {10.1086/167900}, \href {http://adsabs.harvard.edu/abs/1989ApJ...345..245C}
  {345, 245}

\bibitem[\protect\citeauthoryear{{Chauhan}, {Pandey}, {Ogura}, {Jose}, {Ojha},
  {Samal}  \& {Mito}}{{Chauhan} et~al.}{2011}]{Chauhan11}
{Chauhan} N.,  {Pandey} A.~K.,  {Ogura} K.,  {Jose} J.,  {Ojha} D.~K.,  {Samal}
  M.~R.,   {Mito} H.,  2011, \mn@doi [\mnras]
  {10.1111/j.1365-2966.2011.18742.x}, \href
  {http://adsabs.harvard.edu/abs/2011MNRAS.415.1202C} {415, 1202}

\bibitem[\protect\citeauthoryear{{Choi}, {Hachisuka}, {Reid}, {Xu},
  {Brunthaler}, {Menten}  \& {Dame}}{{Choi} et~al.}{2014}]{Choi14}
{Choi} Y.~K.,  {Hachisuka} K.,  {Reid} M.~J.,  {Xu} Y.,  {Brunthaler} A.,
  {Menten} K.~M.,   {Dame} T.~M.,  2014, \mn@doi [\apj]
  {10.1088/0004-637X/790/2/99}, \href
  {http://adsabs.harvard.edu/abs/2014ApJ...790...99C} {790, 99}

\bibitem[\protect\citeauthoryear{{Damineli}, {Almeida}, {Blum}, {Damineli},
  {Navarete}, {Rubinho}  \& {Teodoro}}{{Damineli} et~al.}{2016}]{Damineli16}
{Damineli} A.,  {Almeida} L.~A.,  {Blum} R.~D.,  {Damineli} D.~S.~C.,
  {Navarete} F.,  {Rubinho} M.~S.,   {Teodoro} M.,  2016, \mn@doi [\mnras]
  {10.1093/mnras/stw2122}, \href
  {http://adsabs.harvard.edu/abs/2016MNRAS.463.2653D} {463, 2653}

\bibitem[\protect\citeauthoryear{{Digel}, {Lyder}, {Philbrick}, {Puche}  \&
  {Thaddeus}}{{Digel} et~al.}{1996}]{Digel96}
{Digel} S.~W.,  {Lyder} D.~A.,  {Philbrick} A.~J.,  {Puche} D.,   {Thaddeus}
  P.,  1996, \mn@doi [\apj] {10.1086/176839}, \href
  {http://adsabs.harvard.edu/abs/1996ApJ...458..561D} {458, 561}

\bibitem[\protect\citeauthoryear{{Feigelson} \& {Townsley}}{{Feigelson} \&
  {Townsley}}{2008}]{Feigelson08}
{Feigelson} E.~D.,  {Townsley} L.~K.,  2008, \mn@doi [\apj] {10.1086/524031},
  \href {http://adsabs.harvard.edu/abs/2008ApJ...673..354F} {673, 354}

\bibitem[\protect\citeauthoryear{{Gaia collaboration} et~al.,}{{Gaia
  collaboration} et~al.}{2016}]{Clementini16}
{Gaia collaboration} et~al., 2016, \mn@doi [\aap]
  {10.1051/0004-6361/201629583}, \href
  {http://adsabs.harvard.edu/abs/2016A%26A...595A.133C} {595, A133}

\bibitem[\protect\citeauthoryear{{Gaia collaboration}, {Brown}, {Vallenari}
  et~al.}{{Gaia collaboration} et~al.}{2018}]{Brown18}
{Gaia collaboration} {Brown} A.~G.~A.,  {Vallenari} A.,   et~al., 2018,
  preprint (\mn@eprint {arXiv} {1804.09365})

\bibitem[\protect\citeauthoryear{{Georgelin} \& {Georgelin}}{{Georgelin} \&
  {Georgelin}}{1976}]{Georgelin76}
{Georgelin} Y.~M.,  {Georgelin} Y.~P.,  1976, \aap, \href
  {http://adsabs.harvard.edu/abs/1976A%26A....49...57G} {49, 57}

\bibitem[\protect\citeauthoryear{{Ginsburg}, {Bally}  \& {Williams}}{{Ginsburg}
  et~al.}{2011}]{Ginsburg11}
{Ginsburg} A.,  {Bally} J.,   {Williams} J.~P.,  2011, \mn@doi [\mnras]
  {10.1111/j.1365-2966.2011.19279.x}, \href
  {http://adsabs.harvard.edu/abs/2011MNRAS.418.2121G} {418, 2121}

\bibitem[\protect\citeauthoryear{{Graczyk} et~al.,}{{Graczyk}
  et~al.}{2019}]{Graczyk19}
{Graczyk} D.,  et~al., 2019, arXiv e-prints, \href
  {http://adsabs.harvard.edu/abs/2019arXiv190200589G} {}

\bibitem[\protect\citeauthoryear{{Hachisuka} et~al.,}{{Hachisuka}
  et~al.}{2006}]{Hachisuka06}
{Hachisuka} K.,  et~al., 2006, \mn@doi [\apj] {10.1086/502962}, \href
  {http://adsabs.harvard.edu/abs/2006ApJ...645..337H} {645, 337}

\bibitem[\protect\citeauthoryear{{Hanson}, {Conti}  \& {Rieke}}{{Hanson}
  et~al.}{1996}]{Hanson96}
{Hanson} M.~M.,  {Conti} P.~S.,   {Rieke} M.~J.,  1996, \mn@doi [\apjs]
  {10.1086/192366}, \href {http://adsabs.harvard.edu/abs/1996ApJS..107..281H}
  {107, 281}

\bibitem[\protect\citeauthoryear{{Hanson}, {Kudritzki}, {Kenworthy}, {Puls}  \&
  {Tokunaga}}{{Hanson} et~al.}{2005}]{Hanson05}
{Hanson} M.~M.,  {Kudritzki} R.-P.,  {Kenworthy} M.~A.,  {Puls} J.,
  {Tokunaga} A.~T.,  2005, \mn@doi [\apjs] {10.1086/444363}, \href
  {http://adsabs.harvard.edu/abs/2005ApJS..161..154H} {161, 154}

\bibitem[\protect\citeauthoryear{{Hoag}, {Johnson}, {Iriarte}, {Mitchell},
  {Hallam}  \& {Sharpless}}{{Hoag} et~al.}{1961}]{Hoag61}
{Hoag} A.~A.,  {Johnson} H.~L.,  {Iriarte} B.,  {Mitchell} R.~I.,  {Hallam}
  K.~L.,   {Sharpless} S.,  1961, Publications of the U.S.~Naval Observatory
  Second Series, \href {http://adsabs.harvard.edu/abs/1961PUSNO..17..343H} {17,
  344}

\bibitem[\protect\citeauthoryear{{Hou} \& {Han}}{{Hou} \& {Han}}{2014}]{Hou14}
{Hou} L.~G.,  {Han} J.~L.,  2014, \mn@doi [\aap] {10.1051/0004-6361/201424039},
  \href {http://adsabs.harvard.edu/abs/2014A%26A...569A.125H} {569, A125}

\bibitem[\protect\citeauthoryear{{Humphreys}}{{Humphreys}}{1978}]{Humphreys78}
{Humphreys} R.~M.,  1978, \mn@doi [\apjs] {10.1086/190559}, \href
  {http://adsabs.harvard.edu/abs/1978ApJS...38..309H} {38, 309}

\bibitem[\protect\citeauthoryear{{Imai}, {Kameya}, {Sasao}, {Miyoshi},
  {Deguchi}, {Horiuchi}  \& {Asaki}}{{Imai} et~al.}{2000}]{Imai00}
{Imai} H.,  {Kameya} O.,  {Sasao} T.,  {Miyoshi} M.,  {Deguchi} S.,  {Horiuchi}
  S.,   {Asaki} Y.,  2000, \mn@doi [\apj] {10.1086/309165}, \href
  {http://adsabs.harvard.edu/abs/2000ApJ...538..751I} {538, 751}

\bibitem[\protect\citeauthoryear{{Junqueira}, {L{\'e}pine}, {Braga}  \&
  {Barros}}{{Junqueira} et~al.}{2013}]{Junqueira13}
{Junqueira} T.~C.,  {L{\'e}pine} J.~R.~D.,  {Braga} C.~A.~S.,   {Barros} D.~A.,
   2013, \mn@doi [\aap] {10.1051/0004-6361/201219769}, \href
  {http://adsabs.harvard.edu/abs/2013A%26A...550A..91J} {550, A91}

\bibitem[\protect\citeauthoryear{{Kawamura} \& {Masson}}{{Kawamura} \&
  {Masson}}{1998}]{Kawamura98}
{Kawamura} J.~H.,  {Masson} C.~R.,  1998, \mn@doi [\apj] {10.1086/306472},
  \href {http://adsabs.harvard.edu/abs/1998ApJ...509..270K} {509, 270}

\bibitem[\protect\citeauthoryear{{Kiminki}, {Kim}, {Bagley}, {Sherry}  \&
  {Rieke}}{{Kiminki} et~al.}{2015}]{Kiminki15}
{Kiminki} M.~M.,  {Kim} J.~S.,  {Bagley} M.~B.,  {Sherry} W.~H.,   {Rieke}
  G.~H.,  2015, \mn@doi [\apj] {10.1088/0004-637X/813/1/42}, \href
  {http://adsabs.harvard.edu/abs/2015ApJ...813...42K} {813, 42}

\bibitem[\protect\citeauthoryear{{Kounkel} et~al.,}{{Kounkel}
  et~al.}{2018}]{Kounkel18}
{Kounkel} M.,  et~al., 2018, \mn@doi [\aj] {10.3847/1538-3881/aad1f1}, \href
  {http://adsabs.harvard.edu/abs/2018AJ....156...84K} {156, 84}

\bibitem[\protect\citeauthoryear{{Kudritzki} \& {Puls}}{{Kudritzki} \&
  {Puls}}{2000}]{Kudritzki00}
{Kudritzki} R.-P.,  {Puls} J.,  2000, \mn@doi [\araa]
  {10.1146/annurev.astro.38.1.613}, \href
  {http://adsabs.harvard.edu/abs/2000ARA%26A..38..613K} {38, 613}

\bibitem[\protect\citeauthoryear{{Kwon} \& {Lee}}{{Kwon} \&
  {Lee}}{1983}]{Kwon83}
{Kwon} S.~M.,  {Lee} S.-W.,  1983, Journal of Korean Astronomical Society,
  \href {http://adsabs.harvard.edu/abs/1983JKAS...16....7K} {16, 7}

\bibitem[\protect\citeauthoryear{{Lada}, {Elmegreen}, {Cong}  \&
  {Thaddeus}}{{Lada} et~al.}{1978}]{Lada78}
{Lada} C.~J.,  {Elmegreen} B.~G.,  {Cong} H.-I.,   {Thaddeus} P.,  1978,
  \mn@doi [\apjl] {10.1086/182826}, \href
  {http://adsabs.harvard.edu/abs/1978ApJ...226L..39L} {226, L39}

\bibitem[\protect\citeauthoryear{{Lindegren} et~al.}{{Lindegren}
  et~al.}{2018a}]{LL:LL-124}
{Lindegren} L.,  et~al., 2018a, {R}e-normalising the astrometric chi-square in
  {G}aia {D}{R}2, GAIA-C3-TN-LU-LL-124, \url
  {http://www.rssd.esa.int/doc_fetch.php?id=3757412}

\bibitem[\protect\citeauthoryear{{Lindegren} et~al.,}{{Lindegren}
  et~al.}{2018b}]{Lindegren18}
{Lindegren} L.,  et~al., 2018b, \mn@doi [\aap] {10.1051/0004-6361/201832727},
  \href {http://adsabs.harvard.edu/abs/2018A%26A...616A...2L} {616, A2}

\bibitem[\protect\citeauthoryear{{L{\'o}pez-Corredoira}}{{L{\'o}pez-Corredoira}}{2014}]{Lopez14}
{L{\'o}pez-Corredoira} M.,  2014, \mn@doi [\aap] {10.1051/0004-6361/201423505},
  \href {http://adsabs.harvard.edu/abs/2014A%26A...563A.128L} {563, A128}

\bibitem[\protect\citeauthoryear{{Luri} et~al.,}{{Luri} et~al.}{2018}]{Luri18}
{Luri} X.,  et~al., 2018, preprint, \href
  {http://adsabs.harvard.edu/abs/2018arXiv180409376L} {} (\mn@eprint {arXiv}
  {1804.09376})

\bibitem[\protect\citeauthoryear{{Ma{\'{\i}}z-Apell{\'a}niz}}{{Ma{\'{\i}}z-Apell{\'a}niz}}{2001}]{Apellaniz01}
{Ma{\'{\i}}z-Apell{\'a}niz} J.,  2001, \mn@doi [\aj] {10.1086/320399}, \href
  {http://adsabs.harvard.edu/abs/2001AJ....121.2737M} {121, 2737}

\bibitem[\protect\citeauthoryear{{Ma{\'{\i}}z Apell{\'a}niz}}{{Ma{\'{\i}}z
  Apell{\'a}niz}}{2005}]{Apellaniz05}
{Ma{\'{\i}}z Apell{\'a}niz} J.,  2005, in {Turon} C.,  {O'Flaherty} K.~S.,
  {Perryman} M.~A.~C.,  eds,  ESA Special Publication Vol. 576, The
  Three-Dimensional Universe with Gaia. p.~179 (\mn@eprint {}
  {astro-ph/0411346})

\bibitem[\protect\citeauthoryear{{Massey}, {Johnson}  \&
  {Degioia-Eastwood}}{{Massey} et~al.}{1995}]{Massey95}
{Massey} P.,  {Johnson} K.~E.,   {Degioia-Eastwood} K.,  1995, \mn@doi [\apj]
  {10.1086/176474}, \href {http://adsabs.harvard.edu/abs/1995ApJ...454..151M}
  {454, 151}

\bibitem[\protect\citeauthoryear{{Menten}, {Johnston}, {Wadiak}, {Walmsley}  \&
  {Wilson}}{{Menten} et~al.}{1988}]{Menten88}
{Menten} K.~M.,  {Johnston} K.~J.,  {Wadiak} E.~J.,  {Walmsley} C.~M.,
  {Wilson} T.~L.,  1988, \mn@doi [\apjl] {10.1086/185232}, \href
  {http://adsabs.harvard.edu/abs/1988ApJ...331L..41M} {331, L41}

\bibitem[\protect\citeauthoryear{{Mois{\'e}s}, {Damineli}, {Figuer{\^e}do},
  {Blum}, {Conti}  \& {Barbosa}}{{Mois{\'e}s} et~al.}{2011}]{Moises11}
{Mois{\'e}s} A.~P.,  {Damineli} A.,  {Figuer{\^e}do} E.,  {Blum} R.~D.,
  {Conti} P.~S.,   {Barbosa} C.~L.,  2011, \mn@doi [\mnras]
  {10.1111/j.1365-2966.2010.17713.x}, \href
  {http://adsabs.harvard.edu/abs/2011MNRAS.411..705M} {411, 705}

\bibitem[\protect\citeauthoryear{{Navarete}, {Figueredo}, {Damineli},
  {Mois{\'e}s}, {Blum}  \& {Conti}}{{Navarete} et~al.}{2011}]{Navarete11}
{Navarete} F.,  {Figueredo} E.,  {Damineli} A.,  {Mois{\'e}s} A.~P.,  {Blum}
  R.~D.,   {Conti} P.~S.,  2011, \mn@doi [\aj] {10.1088/0004-6256/142/3/67},
  \href {http://adsabs.harvard.edu/abs/2011AJ....142...67N} {142, 67}

\bibitem[\protect\citeauthoryear{{Oey}, {Watson}, {Kern}  \& {Walth}}{{Oey}
  et~al.}{2005}]{Oey05}
{Oey} M.~S.,  {Watson} A.~M.,  {Kern} K.,   {Walth} G.~L.,  2005, \mn@doi [\aj]
  {10.1086/426333}, \href {http://adsabs.harvard.edu/abs/2005AJ....129..393O}
  {129, 393}

\bibitem[\protect\citeauthoryear{{Ogura} \& {Ishida}}{{Ogura} \&
  {Ishida}}{1976}]{Ogura76}
{Ogura} K.,  {Ishida} K.,  1976, \pasj, \href
  {http://adsabs.harvard.edu/abs/1976PASJ...28..651O} {28, 651}

\bibitem[\protect\citeauthoryear{{Ojha} et~al.,}{{Ojha} et~al.}{2004}]{Ojha04}
{Ojha} D.~K.,  et~al., 2004, \mn@doi [\apj] {10.1086/420876}, \href
  {http://adsabs.harvard.edu/abs/2004ApJ...608..797O} {608, 797}

\bibitem[\protect\citeauthoryear{{Reid} et~al.,}{{Reid} et~al.}{2014}]{Reid14}
{Reid} M.~J.,  et~al., 2014, \mn@doi [\apj] {10.1088/0004-637X/783/2/130},
  \href {http://adsabs.harvard.edu/abs/2014ApJ...783..130R} {783, 130}

\bibitem[\protect\citeauthoryear{{Reid}, {Dame}, {Menten}  \&
  {Brunthaler}}{{Reid} et~al.}{2016}]{Reid16}
{Reid} M.~J.,  {Dame} T.~M.,  {Menten} K.~M.,   {Brunthaler} A.,  2016, \mn@doi
  [\apj] {10.3847/0004-637X/823/2/77}, \href
  {http://adsabs.harvard.edu/abs/2016ApJ...823...77R} {823, 77}

\bibitem[\protect\citeauthoryear{{Riess} et~al.,}{{Riess}
  et~al.}{2018}]{Riess18}
{Riess} A.~G.,  et~al., 2018, \mn@doi [\apj] {10.3847/1538-4357/aac82e}, \href
  {http://adsabs.harvard.edu/abs/2018ApJ...861..126R} {861, 126}

\bibitem[\protect\citeauthoryear{{Roccatagliata}, {Bouwman}, {Henning},
  {Gennaro}, {Feigelson}, {Kim}, {Sicilia-Aguilar}  \&
  {Lawson}}{{Roccatagliata} et~al.}{2011}]{Roccatagliata11}
{Roccatagliata} V.,  {Bouwman} J.,  {Henning} T.,  {Gennaro} M.,  {Feigelson}
  E.,  {Kim} J.~S.,  {Sicilia-Aguilar} A.,   {Lawson} W.~A.,  2011, \mn@doi
  [\apj] {10.1088/0004-637X/733/2/113}, \href
  {http://adsabs.harvard.edu/abs/2011ApJ...733..113R} {733, 113}

\bibitem[\protect\citeauthoryear{{Rom{\'a}n-Z{\'u}{\~n}iga}, {Ybarra},
  {Meg{\'\i}as}, {Tapia}, {Lada}  \& {Alves}}{{Rom{\'a}n-Z{\'u}{\~n}iga}
  et~al.}{2015}]{Roman15}
{Rom{\'a}n-Z{\'u}{\~n}iga} C.~G.,  {Ybarra} J.~E.,  {Meg{\'\i}as} G.~D.,
  {Tapia} M.,  {Lada} E.~A.,   {Alves} J.~F.,  2015, \mn@doi [\aj]
  {10.1088/0004-6256/150/3/80}, \href
  {https://ui.adsabs.harvard.edu/#abs/2015AJ....150...80R} {150, 80}

\bibitem[\protect\citeauthoryear{{Russeil}}{{Russeil}}{2003}]{Russeil03}
{Russeil} D.,  2003, \mn@doi [\aap] {10.1051/0004-6361:20021504}, \href
  {http://adsabs.harvard.edu/abs/2003A%26A...397..133R} {397, 133}

\bibitem[\protect\citeauthoryear{{Sch{\"o}nrich}, {McMillan}  \&
  {Eyer}}{{Sch{\"o}nrich} et~al.}{2019}]{Schoenrich19}
{Sch{\"o}nrich} R.,  {McMillan} P.,   {Eyer} L.,  2019, arXiv e-prints, \href
  {http://adsabs.harvard.edu/abs/2019arXiv190202355S} {}

\bibitem[\protect\citeauthoryear{{Skrutskie} et~al.,}{{Skrutskie}
  et~al.}{2006}]{Skrutskie06}
{Skrutskie} M.~F.,  et~al., 2006, \mn@doi [\aj] {10.1086/498708}, \href
  {http://adsabs.harvard.edu/abs/2006AJ....131.1163S} {131, 1163}

\bibitem[\protect\citeauthoryear{{Stassun} \& {Torres}}{{Stassun} \&
  {Torres}}{2018}]{Stassun18}
{Stassun} K.~G.,  {Torres} G.,  2018, \mn@doi [\apj]
  {10.3847/1538-4357/aacafc}, \href
  {http://adsabs.harvard.edu/abs/2018ApJ...862...61S} {862, 61}

\bibitem[\protect\citeauthoryear{{Stead} \& {Hoare}}{{Stead} \&
  {Hoare}}{2009}]{Stead09}
{Stead} J.~J.,  {Hoare} M.~G.,  2009, \mn@doi [\mnras]
  {10.1111/j.1365-2966.2009.15530.x}, \href
  {http://adsabs.harvard.edu/abs/2009MNRAS.400..731S} {400, 731}

\bibitem[\protect\citeauthoryear{{Sung} et~al.,}{{Sung} et~al.}{2017}]{Sung17}
{Sung} H.,  et~al., 2017, \mn@doi [\apjs] {10.3847/1538-4365/aa6d76}, \href
  {http://adsabs.harvard.edu/abs/2017ApJS..230....3S} {230, 3}

\bibitem[\protect\citeauthoryear{{Urquhart}, {Figura}, {Moore}, {Hoare},
  {Lumsden}, {Mottram}, {Thompson}  \& {Oudmaijer}}{{Urquhart}
  et~al.}{2014}]{Urquhart14}
{Urquhart} J.~S.,  {Figura} C.~C.,  {Moore} T.~J.~T.,  {Hoare} M.~G.,
  {Lumsden} S.~L.,  {Mottram} J.~C.,  {Thompson} M.~A.,   {Oudmaijer} R.~D.,
  2014, \mn@doi [\mnras] {10.1093/mnras/stt2006}, \href
  {http://adsabs.harvard.edu/abs/2014MNRAS.437.1791U} {437, 1791}

\bibitem[\protect\citeauthoryear{{Vasilevskis}, {Sanders}  \& {van
  Altena}}{{Vasilevskis} et~al.}{1965}]{Vasilevskis65}
{Vasilevskis} S.,  {Sanders} W.~L.,   {van Altena} W.~F.,  1965, \mn@doi [\aj]
  {10.1086/109821}, \href {http://adsabs.harvard.edu/abs/1965AJ.....70..806V}
  {70, 806}

\bibitem[\protect\citeauthoryear{{Wenger}, {Balser}, {Anderson}  \&
  {Bania}}{{Wenger} et~al.}{2018}]{Wenger18}
{Wenger} T.~V.,  {Balser} D.~S.,  {Anderson} L.~D.,   {Bania} T.~M.,  2018,
  \mn@doi [\apj] {10.3847/1538-4357/aaaec8}, \href
  {http://adsabs.harvard.edu/abs/2018ApJ...856...52W} {856, 52}

\bibitem[\protect\citeauthoryear{{Xu}, {Reid}, {Zheng}  \& {Menten}}{{Xu}
  et~al.}{2006}]{Xu06}
{Xu} Y.,  {Reid} M.~J.,  {Zheng} X.~W.,   {Menten} K.~M.,  2006, \mn@doi
  [Science] {10.1126/science.1120914}, \href
  {http://adsabs.harvard.edu/abs/2006Sci...311...54X} {311, 54}

\bibitem[\protect\citeauthoryear{{Zhou}, {Yang}, {Fang}, {Su}, {Sun}  \&
  {Chen}}{{Zhou} et~al.}{2016}]{Zhou16}
{Zhou} X.,  {Yang} J.,  {Fang} M.,  {Su} Y.,  {Sun} Y.,   {Chen} Y.,  2016,
  \mn@doi [\apj] {10.3847/0004-637X/833/1/4}, \href
  {https://ui.adsabs.harvard.edu/#abs/2016ApJ...833....4Z} {833, 4}

\bibitem[\protect\citeauthoryear{{Zinn}, {Pinsonneault}, {Huber}  \&
  {Stello}}{{Zinn} et~al.}{2018}]{Zinn18}
{Zinn} J.~C.,  {Pinsonneault} M.~H.,  {Huber} D.,   {Stello} D.,  2018, arXiv
  e-prints, \href {http://adsabs.harvard.edu/abs/2018arXiv180502650Z} {}

\makeatother
\end{thebibliography}
